\documentclass[fleqn,usenatbib]{mnras}

\usepackage{newtxtext,newtxmath}

\usepackage[T1]{fontenc}

\DeclareRobustCommand{\VAN}[3]{#2}
\let\VANthebibliography\thebibliography
\def\thebibliography{\DeclareRobustCommand{\VAN}[3]{##3}\VANthebibliography}

\usepackage{graphicx}	
\usepackage{amsmath}	
\usepackage{float}
\usepackage{subcaption}

\title[An Analysis of M-Dwarfs using \textit{CHEOPS}]{The EBLM project - VIII. First results for M-dwarf mass, radius and effective temperature measurements using \textit{CHEOPS} light curves\thanks{Based in part on observations collected at the Observatoire de Haute-Provence (CNRS), France.
}}
\author[M. I. Swayne et al.]{
M. I. Swayne,$^{1}$\thanks{E-mail: m.i.swayne@keele.ac.uk}
P. F. L. Maxted,$^{1}$ 
A. H. M. J. Triaud,$^{2}$ 
S. G. Sousa,$^{3}$ 
C. Broeg,$^{4,5}$ 
H.-G. Flor{\'e}n,$^{6}$ 
\newauthor P. Guterman,$^{7,8}$ 
A. E. Simon,$^{4}$ 
I. Boisse,$^{7}$ 
A. Bonfanti,$^{9}$ 
D. Martin,$^{10}$\thanks{NASA Sagan Fellow} 
A. Santerne,$^{7}$ 
S. Salmon,$^{11}$ 
\newauthor M. R. Standing,$^{2}$ 
V. Van Grootel,$^{12}$ 
T. G. Wilson,$^{13}$ 
Y. Alibert,$^{4}$ 
R. Alonso,$^{14,15}$ 
G. Anglada Escudé,$^{16,17}$ 
\newauthor J. Asquier,$^{18}$
T. B{\'a}rczy,$^{19}$
D. Barrado,$^{20}$ 
S. C. C. Barros,$^{3,21}$ 
M. Battley,$^{22,23}$ 
W. Baumjohann,$^{9}$ 
M. Beck,$^{11}$ 
\newauthor T. Beck,$^{4}$
A. Bekkelien,$^{11}$ 
W. Benz,$^{4,5}$ 
N. Billot,$^{11}$ 
X. Bonfils,$^{24}$ 
A. Brandeker,$^{6}$ 
M.-D. Busch,$^{4,5}$ 
\newauthor J. Cabrera,$^{25}$ 
S. Charnoz,$^{26}$ 
A. Collier Cameron,$^{13}$ 
Sz. Csizmadia,$^{25}$ 
M. B. Davies,$^{27}$ 
M. Deleuil,$^{7}$ 
\newauthor A. Deline,$^{11}$ 
L. Delrez,$^{28,12}$ 
O. D. S. Demangeon,$^{3,21}$ 
B.-O. Demory,$^{5}$ 
G. Dransfield,$^{2}$ 
D. Ehrenreich,$^{11}$ 
\newauthor A. Erikson,$^{25}$ 
A. Fortier,$^{4,5}$ 
L. Fossati,$^{9}$ 
M. Fridlund,$^{29,30}$ 
D. Futyan,$^{11}$ 
D. Gandolfi,$^{31}$ 
M. Gillon,$^{28}$ 
\newauthor M. Guedel,$^{32}$
G. H{\'e}brard,$^{33,34}$
N. Heidari,$^{35,36,7}$ 
C. Hellier,$^{1}$ 
K. Heng,$^{5,22}$
M. Hobson,$^{37,38}$
S. Hoyer,$^{7}$ 
\newauthor K. G. Isaak,$^{39}$ 
L. Kiss,$^{40,41,42}$ 
V. Kunovac Hod\v{z}i\'{c},$^{2}$
S. Lalitha,$^{2}$ 
J. Laskar,$^{43}$ 
A. Lecavelier des Etangs,$^{33}$ 
\newauthor M. Lendl,$^{11}$ 
C. Lovis,$^{11}$ 
D. Magrin,$^{44}$ 
L. Marafatto,$^{44}$ 
J. McCormac,$^{22}$ 
N. Miller,$^{1}$ 
V. Nascimbeni,$^{44}$ 
\newauthor G. Olofsson,$^{6}$ 
R. Ottensamer,$^{32}$ 
I. Pagano,$^{45}$ 
E. Pall{\'e},$^{14,15}$ 
G. Peter,$^{46}$ 
G. Piotto,$^{44,47}$ 
D. Pollacco,$^{22}$ 
\newauthor D. Queloz,$^{11,48}$ 
R. Ragazzoni,$^{44,47}$ 
N. Rando,$^{18}$ 
H. Rauer,$^{25,49,50}$ 
I. Ribas,$^{16,17}$ 
N. C. Santos,$^{3,21}$ 
\newauthor G. Scandariato,$^{45}$
D. Ségransan,$^{11}$
A. M. S. Smith,$^{25}$ 
M. Steinberger,$^{9}$ 
M. Steller,$^{9}$ 
Gy. M. Szab{\'o},$^{51,52}$ 
\newauthor N. Thomas,$^{4}$ 
S. Udry,$^{11}$ 
I. Walter,$^{46}$ 
N. A. Walton,$^{53}$ 
and E. Willett$^{2}$
\\
(Affiliations listed after the references)
}

\date{Accepted XXX. Received YYY; in original form ZZZ}

\pubyear{2021}

\begin{document}
\label{firstpage}
\pagerange{\pageref{firstpage}--\pageref{lastpage}}
\maketitle

\begin{abstract}
The accuracy of theoretical mass, radius and effective temperature values for M-dwarf stars is  an active topic of debate.
Differences between observed and theoretical values have raised the possibility that current theoretical stellar structure and evolution models are inaccurate towards the low-mass end of the main sequence.
To explore this issue we use the \textit{CHEOPS} satellite to obtain high-precision light curves of eclipsing binaries with low mass stellar companions.
We use these light curves combined with the spectroscopic orbit for the solar-type companion to measure the mass, radius and effective temperature of the M-dwarf star.
Here we present the analysis of three eclipsing binaries.
We use the {\fontfamily{qcr}\selectfont pycheops} data analysis software to fit the observed transit and eclipse events of each system.
Two of our systems were also observed by the \textit{TESS} satellite-- we similarly analyse these light curves for comparison.
We find consistent results between \textit{CHEOPS} and \textit{TESS}, presenting three stellar radii and two stellar effective temperature values of low-mass stellar objects.
These initial results from our on-going observing programme with \textit{CHEOPS} show that we can expect to have $\sim$24 new mass, radius and effective temperature measurements for very low mass stars within the next few years.
\end{abstract}




\section{Introduction}

    Understanding the host star is one of the most crucial parts of exoplanet characterisation.
    Exoplanets are mostly observed and analysed through how they effect the stellar signal, such as with the transit and radial velocity methods \citep{santos2020detection}.
    A more accurate measurement of host size and mass thus leads to more accurate values of planetary size and mass.
    The host star's properties are most commonly obtained by finding the closest fit between observable star properties and stellar evolution models \citep[e.g.][]{Baraffe98,Dotter}.
    Therefore, any uncertainties in these models leads to systematic errors in the inferred stellar and exoplanetary properties.
    This has become a potential issue regarding low-mass star systems' recent popularity as targets for exoplanet observation \citep{charbonneau2007dynamics, quirrenbach2014carmenes,gillon2017seven,delrez}.
    Low-mass star systems suffer from a lack of data compared to more massive stars because they are intrinsically much fainter and, hence, harder to study.
    There is a shortfall in direct and precise mass and radius measurements of these systems, with effective temperature measurements being rarer still.
    The EBLM project \citep{triaud} was launched to address this lack of fundamental data for M-dwarfs. 
    Hundreds of eclipsing binaries with low-mass companions have been identified using data from the WASP project \citep{pollacco2006wasp}, and we have measured the spectroscopic orbits for the primary stars in more than 100 of these EBLM systems \citep{triaud2017eblm}.
    These data are used to select targets for further study to address lack of precise mass, radius and temperature measurements for low-mass stars, especially below 0.3 solar masses.
    
    A number of studies have reported inconsistencies between the observed radii and M-dwarfs and theoretically predicted radii from models of low-mass stars, an effect commonly called radius inflation \citep[e.g.][]{casagrande2008,torres2010,spada2013,kesseli2018}. 
    Typically, the measured radii are larger than the predicted values for stars of a given mass by a few percent \citep[e.g.][]{morales2009absolute}. 
    There is also a tendency for M-dwarfs to be cooler than predicted by models, such that the luminosity of the star is approximately correct.
    It is currently unclear to what extent radius inflation is due to problems with stellar models, or is the result of bias in the observed radius estimates.
    Possible sources of error from the models involve uncertainties in the input physics of the model, its initial chemical composition and in convection efficiency \citep{tognelli2018theoretical,fernandes2019evolutionary}.
    These would in turn provide an uncertainty to predicted radius.
    It is also possible that some models are missing some physical process that affects the stellar radius.
    The presence of a strong magnetic field or magnetic activity could inhibit the convective energy transport present in lower-mass stars \citep{chabrier2007evolution}.
    This could result in the inflation effect as the star attempts to maintain a constant energy flux through the surface.
    Rotation in eclipsing binaries has also been proposed as a potential cause.
    Tidal interactions between the two bodies in the system could increase the speed of the internal stellar dynamo leading to increased activity \citep{ribas2006}.
    Radius inflation therefore could be an observational bias caused by using eclipsing binaries to obtain radii from M-dwarfs.
    However cases of long-period eclipsing binaries \citep{irwin2011lspm} and isolated M-dwarfs \citep{spada2013, van2018stellar} showing similar inflation to short-period eclipsing binaries casts doubt on tidal interactions being the sole cause .
    The effect of metallicity on very low mass stars is also debated as a possible cause for inflation with its effect on the opacity in the outer layers of the star.
    In their revision of the age of CM Dra, \cite{feiden2014revised} find a reduction of observed mass-radius discrepancies from 6\% to 2\% upon obtaining more accurate metallicity and age measurements for this binary star.
    Metallicity measurements for EBLM systems are more reliable than M+M binaries like CM Dra because the spectrum of a solar-type star is much less complex and crowded than the spectrum of a rapidly-rotating M-dwarf star. 
    Radius measurements for several EBLM systems by \cite{von2019eblm} suggest that the metallicity may have a measurable effect on stellar radius.
    Therefore the accuracy of metallicity values is important when considering the radius inflation problem.
    Large uncertainties in metallicity, such as those in the order of 0.2 dex as seen in \cite{olander2021comparative}, could lead to differences in radius residuals of $\sim$0.024 according to the metallicity dependent relation described in \cite{von2019eblm}.
    Finally, there has been recent disagreement on the reality of the effect.
    \cite{parsons2018scatter} reports that 75\% of their objects are up to 12\% inflated.
    However, two papers in the EBLM project \citep{von2019eblm,gill2019eblm} find little evidence of inflation in their samples of 10 and 5 objects respectively.
    A much larger sample of precise and accurate mass, radius and effective measurements for M-dwarfs of known metallicity is needed so that we can reliably estimate the properties of low-mass host stars in planetary systems.

    The \textit{CHEOPS} mission \citep{Benz} is the first small (S-class) European Space Agency mission.
    Launched on the 18th of December 2019, it has been designed primarily to perform ultrahigh-precision photometry of bright stars that are known to host exoplanet systems.
    The \textit{CHEOPS} guaranteed-time observing programme includes a small number of ``Ancillary Science'' programmes  where the stars observed do not host exoplanets, but where the observations made are relevant to exoplanet science.
    This includes our programme to use the capabilities of \textit{CHEOPS} to explore the radius inflation problem.
    Additionally, in measurements of M-dwarf effective temperature in EBLM systems, there is the possibility of some unrealised systematic error, with different studies reporting widely different results for the same object \citep[e.g.][]{Yilen,swayne2020tess}.
    Through obtaining high precision observations of secondary eclipses we can compare to previous observations and explore any potential systematic effect. 

    In this paper we present our  analysis of the first three targets in our \textit{CHEOPS} observing programme with a complete set of observations -- EBLM~J1741+31, EBLM~J1934$-$42 and EBLM~J2046+06.
    EBLM~J1741+31 and EBLM~J1934$-$42 have also been observed by the \textit{TESS} satellite \citep{Ricker}.
    This gives us an opportunity to test the reliability of our methods to measure mass, radius and effective temperature by comparing the results from the two instruments.
    Our observations, data reduction and methods to characterise the host star are outlined in Section~\ref{sec:obs}.
    The analysis of the light curves and results are described in Section~\ref{sec:analysis}.
    We discuss our results in the context of previous mass, radius and effective temperature measurements for M-dwarfs in Section~\ref{sec:discuss}, and give our conclusions as to the future prospects for our observing programme in Section~\ref{sec:conc}.
    
\begin{table*}
      \caption{A log of observation dates and details for each target visit. Sp. Type is the estimated spectral type of the primary star. Effic. is the fraction of the observing interval covered by valid observations of the target. $R_{\rm ap}$ is the aperture radius used to compute the light curve analysed in this paper.}
         \label{ObsLog}
    $$ 
         \begin{tabular}{lcrrrrrrrr}
            \hline\hline
            \noalign{\smallskip}
            Event & \multicolumn{1}{c}{Target}& \multicolumn{1}{l}{V} & \multicolumn{1}{l}{Start Date} & \multicolumn{1}{l}{Duration} & \multicolumn{1}{l}{$T_\mathrm{exp}$} & \multicolumn{1}{l}{Effic.} & \multicolumn{1}{l}{File key} & \multicolumn{1}{l}{$R_{\rm ap}$}\\
             &  \multicolumn{1}{c}{Sp. Type}&\multicolumn{1}{l}{(mag)} & \multicolumn{1}{l}{(UTC)} & \multicolumn{1}{l}{[s]} & \multicolumn{1}{l}{[s]}  & (\%) &      & [pixels]    \\
            \noalign{\smallskip}
            \hline
            \noalign{\smallskip}
            Transit & J1741+31 & 11.7 & 2020-06-13T08:20:00 & 27794 & 60 & 67.8 & CH\_PR100037\_TG014601\_V0102 & 30.0    \\
            Eclipse$^{\dagger}$ & G0V& & 2020-06-10T08:12:58 & 29098 & 60 & 63.0 & CH\_PR100037\_TG014501\_V0102 & 30.0  \\
            Transit & J1934-42 & 12.62 & 2020-06-27T13:43:57 & 28387 & 60 & 60.7 & CH\_PR100037\_TG015001\_V0100 & 25.0 \\
            Eclipse &G8V & & 2020-07-13T09:47:00 & 28387 & 60 & 61.1 & CH\_PR100037\_TG014901\_V0100 & 25.0   \\
            Transit & J2046+06 & 9.86 & 2020-08-28T22:08:00 & 35676 & 60 & 81.1 & CH\_PR100037\_TG015601\_V0100 & 25.0   \\
            Eclipse &F8V & & 2020-07-03T11:34:00 & 42313 & 60 & 66.7 & CH\_PR100037\_TG015501\_V0100 & 25.0  \\
            \noalign{\smallskip}
            \hline
            \multicolumn{9}{@{}l}{$^{\dagger}$ Does not cover the phase of superior conjunction.}
         \end{tabular}
    $$
\end{table*}

\section{Observations and methods}
\label{sec:obs}

     Our three targets are all detached eclipsing binary stars in which a solar-type star is eclipsed by an M-dwarf. 
     The log of our observations is given in Table \ref{ObsLog}.
     The observations were made as part of the \textit{CHEOPS} Guaranteed Time Observation (GTO) programme ID-037: Eclipsing binaries with very low mass stars.
     This programme seeks to observe primary and secondary eclipses of 25 EBLM systems.
     \textit{CHEOPS} observes stars from low-Earth orbit, so observations are interrupted by occultation of the target by the Earth and passages through the South Atlantic Anomaly.
     These gaps in the light curve can be up to 44 and 19 minutes, respectively. 
 
    The raw data were processed using version 13 of the \textit{CHEOPS} data reduction pipeline \citep[DRP,][]{hoyer2020expected}.
    The DRP performs image correction for environmental and instrumental effects before performing aperture photometry of the target.
    As explained in \cite{hoyer2020expected}, the Gaia DR2 catalogue \citep{gaia2018gaia} is used by the DRP to simulate an observations' field of view (FoV) in order to estimate the level of contamination present in the photometric aperture.
    The DRP also accounts for the rotating FoV of \textit{CHEOPS}, where other stars in the image can create ``smear'' trails and contaminate the photometric aperture.
    The smear effect is corrected by the DRP while the contamination produced by nearby stars is recorded in the DRP data products, allowing the user to include or ignore the contamination correction provided.
    The final photometry is extracted by the DRP using three different fixed aperture sizes labelled ``RINF'', ``DEFAULT'' and ``RSUP'' (at radii of 22.5, 25.0 and 30.0 pixels, respectively) and a further ``OPTIMAL'' aperture whose size is dependent upon the FoV contamination.
    The observed and processed data are made available on the Data Analysis Center for Exoplanets (DACE) web platform\footnote{The DACE platform is available at \url{http://dace.unige.ch}}.
    We downloaded our data from DACE using {\fontfamily{qcr}\selectfont pycheops}\footnote{\url{https://pypi.org/project/pycheops/}}, a \textsc{python} module developed for the analysis of data from the \textit{CHEOPS} mission \citep{pycheops}.
    We fitted the light curves from all four apertures and found that different choice of aperture radius has a negligible impact on the results.
    Therefore, for EBLM~J1741+31 and EBLM~J1934$-$42, we selected the aperture radius that gave the minimum median absolute deviation (MAD) of the point-to-point differences in the light curve of the eclipse visit. 
    We then used the chosen aperture type for the respective transit visits.
    For EBLM~J2046+06 this criterion resulted in slightly different aperture radii for the two visits from the preferred OPTIMAL aperture (25.5 and 26.0 pixels), so we used the DEFAULT aperture instead. 
    
    The \textit{TESS} survey is split into overlapping $90^{\circ} \times 24^{\circ}$ degree sky sectors over both northern and southern hemispheres with each sector being observed for approximately one month.
    EBLM~J1741+31 (TIC 18319090) was observed in Sectors 25 and 26 of the survey as part of the Guest Investigator programs G022156 and G022253, with 2-minute cadence data made available.
    EBLM~J1934$-$42 (TIC 143291764) was observed in Sectors 13 and 27 of the survey as part of the Guest Investigator programs G011278 and G03216, with 2-minute cadence data made available.
    Data was reduced by the Science Processing Operations Center Pipeline \citep[SPOC;][]{jenkins2016tess} and made available from the Mikulski Archive for Space Telescopes (MAST)\footnote{\url{https://mast.stsci.edu}} web service.
    We used the Pre-search Data Conditioned Simple Aperture Photometry (PDCSAP) flux data for our analysis.
    Any cadences in the light curve with severe quality issues were ignored using the ``default'' bitmask 175 \citep{tenenbaum}.
    The TESS light curve of EBLM~J1741+31 shows a smooth variation with an amplitude $\sim 0.2$\% in the flux between the transits. 
    To remove this variability we divided the light curve by a low-order polynomial fitted by least-squares to the data between the transits. 
    EBLM~J1934$-$42 shows variability in the TESS light curve with an amplitude of about 1\% on timescales of a few days. 
    This may be due to moderate stellar activity modulated by stellar rotation.
    To remove this low-frequency noise we fit the data between the transits with a Gaussian process (GP) calculated using the {\fontfamily{qcr}\selectfont celerite} \citep{celerite} software package. 
    The kernel of the GP is the stochastically-driven damped simple harmonic oscillator function defined by \citeauthor{celerite}. 
    We then divide the entire light curve by the GP predicted by the best-fit hyper-parameters.
    
    The spectroscopic stellar parameters ($T_{\mathrm{eff}}$, $\log g$, microturbulence ($\xi_{\mathrm{t}}$), [Fe/H]) and respective uncertainties were estimated by using ARES+MOOG, following the same methodology as described in \citet[][]{Sousa-14, Santos-13}.
    For this we used the combined spectra from the individual observations done with SOPHIE for EBLM~J1741+31 and with HARPS observations from ESO programme 1101.C-0721 for EBLM~J1934$-$42 and EBLM~J2046+06.
    For  EBLM~J1741+31 there were 13 individual observations with SOPHIE, with a signal-to-noise ratio (SNR) of 20-50. The combined spectrum has a total SNR $\sim$140.
    For EBLM~J1934$-$42 there were 24 individual observations, with SNR varying between 15-20. The combined spectra has a total SNR $\sim$100.
    For EBLM~J2046+06 there were 22 individual observations, with SNR varying between 50-80.
    The combined spectra has a total SNR $\sim$300.
    We used the ARES code\footnote{The last version of ARES code (ARES v2) can be downloaded at \url{https://github.com/sousasag/ARES}} \citep{Sousa-07, Sousa-15} to measure equivalent widths (EW) of iron lines measured using the list of lines presented in \citet[][]{Sousa-08}.
    A minimization process assuming ionization and excitation equilibrium is used to find convergence for the best set of spectroscopic parameters.
    In this process we use a grid of Kurucz model atmospheres \citep{Kurucz-93} and the radiative transfer code MOOG \citep{Sneden-73}.

    The radii of the three targets was determined using an adapted infrared flux method (IRFM; \citealt{Blackwell1977}) via relationships between the bolometric flux, the stellar angular diameter, the effective temperature, and the parallax, recently detailed in \citet{Schanche2020}.
    For each target, and using a MCMC approach, we built spectral energy distributions (SEDs) from the {\sc atlas} Catalogues \citep{Castelli2003} using the stellar spectral parameters derived above as priors.
    Subsequently, we conducted synthetic photometry by convolving the SEDs with the throughput of the selected photometric bandpasses and compared the resulting fluxes with the observed fluxes in these bandpasses; {\it Gaia} G, G$_{\rm BP}$, and G$_{\rm RP}$, 2MASS J, H, and K, and {\it WISE} W1 and W2 \citep{Skrutskie2006,Wright2010,GaiaCollaboration2020} to obtain the stellar bolometric fluxes and hence the angular diameters.
    These diameters were combined with offset-corrected {\it Gaia} EDR3 parallax \citep{Lindegren2020} to produce the stellar radii given in Table~\ref{ParamsStar}.
    
    The stellar mass $M_{\star}$ and age $t_{\star}$ were inferred from two different stellar evolutionary models, namely the PARSEC\footnote{Padova and Trieste Stellar Evolutionary Code\\ {\tt \url{http://stev.oapd.inaf.it/cgi-bin/cmd}}.} v1.2S code \citep{marigo17} and the CLES code \citep[Code Liègeois d'Évolution Stellaire;][]{scuflaire08}.
    We adopted the stellar effective temperature $T_{\mathrm{eff}}$, metallicity [Fe/H], and radius $R_{\mathrm{IRFM},\star}$ as input parameters and carried out two independent analyses.
    The first analysis used the Isochrone placement algorithm \citep{bonfanti15,bonfanti16} which retrieves the best estimates for both mass and age by interpolating within pre-computed PARSEC grids of isochrones and tracks. The second analysis, instead, returned the mass and age values by directly fitting the input parameters to the CLES models following a Levenberg-Marquadt minimisation \citep{salmon21}.
    Finally, we combined the two different mass and age values to obtain the definitive $M_{\star}$ and $t_{\star}$ parameters; further details can be found in \citet{bonfanti21}.
    The masses obtained are given in Table~\ref{ParamsStar}.
    
    The semi-amplitude of the primary star's spectroscopic orbit, $K$, is required for the calculations of secondary star's mass.
    For EBLM~J1934$-$42 and EBLM~J2046+06 we used values of $K$ from the Binaries Escorted By Orbiting Planets survey \citep[BEBOP,][]{martin2019bebop}.
    For J1741-31 we calculated $K$ from a fit to radial velocity data from the SOPHIE high-resolution \'echelle spectrograph \citep{Perruchot08} mounted on the 193cm telescope at the Observatoire de Haute-Provence (France). 
    Twenty measurements were collected between the dates of 2019-02-24 and 2020-09-03 with a typical exposure time of 1800s leading to a mean uncertainty of $13.7\,{\rm m\,s^{-1}}$.
    These were obtained as part of a Large Programme aiming to detect circumbinary planets \citep[e.g.][]{martin2019bebop}.
    The 20 spectra were obtained in High-Efficiency mode, where the resolution is reduced to 40,000 for a $2.5\times$ gain in throughput over the High-Resolution mode of 75,000.
    All observations were performed with a fibre on the science target and a fibre on the sky.
    The latter is used to remove background contamination originating from the Moon.
    All science and sky spectra were reduced using the SOPHIE Data Reduction Software (DRS) and cross-correlated with a G2 mask to obtain radial-velocities.
    These methods are described in \citet{Baranne96}, and \citet{Courcol15}, and have been shown to produce precisions and accuracies of a few meters per seconds \citep[e.g.][]{Bouchy2013,Hara20}, well below what we typically obtained on this system.
    We used the \textsc{python} module {\fontfamily{qcr}\selectfont ellc} \citep{maxted2016ellc} to model radial velocity.
    In our fit of the Keplerian orbit we accounted for jitter by applying a weight in our log-likelihood function.
    We used the \textsc{python} module {\fontfamily{qcr}\selectfont emcee} \citep{Foreman} to sample the posterior probability distribution of our model parameters.
    The stellar properties and obtained value of $K$ are all displayed in Table \ref{ParamsStar}.
    
\begin{table}
\centering
      \caption{The observed stellar properties of the primary star of our binary targets. 
      Right ascension (RA) and declination (Dec) are coordinates with equinox J2000.0.}
         \label{ParamsStar}
    
         \resizebox{\columnwidth}{!}{\begin{tabular}{lrrr}
            \hline
            \noalign{\smallskip}
                  & \multicolumn{1}{l}{EBLM J1741+31} & \multicolumn{1}{l}{EBLM J1934$-$42} & \multicolumn{1}{l}{EBLM J2046+06}\\
            \noalign{\smallskip}
            \hline
            \noalign{\smallskip}
            Name & TYC 2606$-$1838$-$1 & TIC 143291764 & TYC 524$-$2528$-$1 \\
            RA & 17\:41\:21.27 & 19\:34\:25.69 & 20\:46\:43.88 \\
            Dec & +31\:24\:55.3 & -42\:23\:11.6  &  +06\:18\:09.7 \\
            G (mag) & 11.40 &  11.42 &  9.83 \\
            ${\rm T}_{\rm eff,1}$ ({\rm K}) & $6376 \pm 72$ & $5648 \pm 68$ & $6302 \pm 70$      \\
            $\log g_1$ (cgs) & $4.63 \pm 0.11$ & $4.33 \pm 0.12$ & $3.98 \pm 0.11$   \\
            $\xi_{\mathrm{t}}$ (km/s) & $1.25	\pm 0.05$ & $1.10 \pm 0.04$ & $1.61 \pm 0.05$ \\
            $[\text{Fe/H}]$ & $0.09 \pm 0.05$ & $0.29 \pm 0.05$ & $0.00 \pm 0.05$ \\
            ${\rm R}_{\rm 1} ({\rm R_\odot})$ & $1.336 \pm 0.015$ & $0.996 \pm 0.008$ & $ 1.722 \pm 0.015$   \\
            ${\rm M}_{\rm 1} ({\rm M_\odot})$ & $ 1.270 \pm 0.043$ & $ 1.046 \pm 0.049$ & $ 1.339 \pm 0.056 $  \\
            Age (Gyr.) & $ 1.2 \pm 0.7$ & $ 1.8 \pm 2.0$ & $ 2.8 \pm 0.6$ \\
            $K$ (km/s) & $37.14 \pm 0.04$ & $18.62 \pm 0.01$ & $15.55 \pm 0.01$  \\
            \noalign{\smallskip}
            \hline
         \end{tabular}}
    
\end{table}

\section{Analysis}
\label{sec:analysis}

\begin{table}
      \caption{The priors set for each target during the MultiVisit analysis.}
         \label{Priors}
    $$ 
         \begin{tabular}{lrrr}
            \hline
            \noalign{\smallskip}
            \text{}      & \multicolumn{1}{l}{J1741+31} & \multicolumn{1}{l}{J1934-42} & \multicolumn{1}{l}{J2046+06}\\
            \noalign{\smallskip}
            \hline
            \noalign{\smallskip}
            $f_c$ & $0.3003 \pm 0.0016$ & -- & $-0.1901 \pm 0.0008$  \\
            $f_s$ & $0.4591 \pm 0.0012$ & -- & $0.5545 \pm 0.0004$   \\
            $h_1$ & $0.771 \pm 0.012$ & $0.729 \pm 0.011$ & -- \\
            $h_2$ & $0.420 \pm 0.050$ & $0.398 \pm 0.050$ & -- \\
            $L$ & $0.007 \pm 0.004$ & -- & -- \\
            $\log \rho/\rho_{\odot}$ & $-0.274 \pm 0.021$ & -- & -- \\
            \noalign{\smallskip}
            \hline
         \end{tabular}
    $$
\end{table}

    We analyse the \textit{CHEOPS} light curves for each star in two steps. 
    In the first step, we analyse each \textit{CHEOPS} visit in order to determine initial values for our model parameters, and to determine which nuisance parameters must be included in the model to deal with instrumental noise.
    In the second step, we analyse all the data for each star in a single  Markov chain Monte Carlo (MCMC) analysis to obtain our final results.
    These results are then compared to a MCMC analysis of \textit{TESS} data when available.
    The output from the light curve analysis is then combined with as estimate of for the mass of the primary star and $K$ to determine the mass and radius of the M-dwarf.
    The depth of the secondary eclipse is used together with model spectral energy distributions to estimate the effective temperature of the M-dwarf.

\subsection{CHEOPS visit-by-visit analysis}

    To create the models needed for light curve fitting we used {\fontfamily{qcr}\selectfont pycheops}.
    The transit model uses the qpower2 algorithm \citep{maxted2019q} to calculate the transit light curve assuming a power-2 limb darkening law.
    The parameters used in the model are: the time of mid-primary eclipse $T_0$; the transit depth $D = k^2 = R_2^2/R_1^2$, where $R_2$ and $R_1$ are the radii of the secondary and primary stars;  $b = a \cos{i}/R_1$, where $i$ is the orbital inclination and $a$ is the semimajor axis;  $W = \sqrt{(1+k)^2 - b^2} R_1/(\pi a)$; the eccentricity and argument of periastron dependent parameters $f_s = \sqrt{e} \sin{(\omega)}$ and $f_c = \sqrt{e} \cos{(\omega)}$; the eclipse depth $L$ and the limb-darkening parameters $h_1$ and $h_2$ as defined by \cite{maxted2018}.
    For an eclipsing binary with a circular orbit, $D$, $W$ and $b$ are the depth, width (in phase units) and impact parameter of the eclipse, respectively. 
    For each target we obtained one primary and one secondary eclipse so the orbital period, $P$, has to be fixed at a known value.
    For EBLM~J1741+31 and EBLM~J1934$-$42 we fixed $P$ to the value obtained from our analysis of the \textit{TESS} light curve.
    For EBLM~J2046+06 we fixed the orbital period at the value reported by \cite{martin2019bebop}. 
    To better constrain our fit, Gaussian priors were put on $f_c$ and $f_s$ using $e$ and $\omega$ measurements from the spectroscopic orbit.
    The orbital eccentricity of EBLM~J1934$-$42 is very small so we assumed a circular orbit for our analysis.
    For EBLM~J1741+31 and EBLM~J1934$-$42, which have partial eclipses, the eclipses do not constrain the limb darkening properties of the star so we place Gaussian priors on $h_1$ and $h_2$.
    These priors are listed in Table \ref{Priors}.
    The values of  $h_1$ and $h_2$ appropriate for the values of [Fe/H], ${\rm T}_{\rm eff,1}$ and $\log g$ given in Section~\ref{sec:obs} are found using interpolation in the  data tables presented in \cite{maxted2018} based on the limb-darkening profiles from the STAGGER-grid \citep{magic2015stagger}.
    An offset (0.01 for $h_1$, $-$0.045 for $h_2$) was then applied based on the offset between empirical and tabulated values of these limb darkening parameters observed in the \textit{Kepler} bandpass by \citet{maxted2018}.
    
    \textit{CHEOPS} light curves can be affected by trends correlated with satellite roll angle, the varying contamination of the photometric aperture, the background level in the images, and the estimated correction for smear trails from nearby stars.
    These trends are modelled using linear decorrelation against these parameters or, for roll angle $\phi$, $\sin(\phi)$, $\cos(\phi)$, $\sin(2 phi)$, etc.
    The  coefficients for each trend are optimised simultaneously with the parameters of the transit or eclipse model in a least squares fit to all the data in each visit.
    In the case of the eclipse events, fits to individual visits were performed with all orbital parameters apart from eclipse depth fixed at the parameters derived from the fit to the transit.
    To select decorrelation parameters for each visit we did an initial fit to each light curve with no decorrelation and used the RMS of the residuals from this fit, $\sigma_p$, to set a normal prior on the decorrelation parameters, $\mathcal{N}(0, \sigma_p)$ or, for ${df}/{dt}$,  $\mathcal{N}(0, \sigma_p/\Delta t)$ where $\Delta t$ is the duration of the visit.
    We then added decorrelation parameters to the fit one-by-one, selecting the parameter with the lowest Bayes factor $B_p = e^{-(p/\sigma_p)^2/2}\,\sigma_0/\sigma_p$ at each step, where $\sigma_0$ is the standard error on the decorrelation parameter from the least-squares fit \citep{pycheops}.
    We stop adding decorrelation parameters when $B_p>1$ for all remaining parameters.
    This process sometimes leads to a set of parameters including some that are strongly correlated with one another and so are therefore not well determined, i.e. they have large Bayes factors.
    We therefore go through a process of repeatedly removing the parameter with the largest Bayes factor if any of the parameters have a Bayes factors $B_p>1$.
    The second step of this process typically removes no more than 1 or 2 parameters.

\subsection{CHEOPS MultiVisit analysis}

    We used the \textsc{MultiVisit} function in {\fontfamily{qcr}\selectfont pycheops} to do a combined analysis of both visits for each target.
    Decorrelation against trends with roll angle were done implicitly using the method described in \citep{pycheops}, i.e. by modifying the calculation of the likelihood to account for the decorrelation against roll angle without explicitly calculating the nuisance parameters  $df/d\sin(\phi)$, $df/d\cos(\phi)$, etc. 
    The same Gaussian priors for $f_c$ and $f_s$, $h_1$ and $h_2$ were used as for the analysis of individual visits.
    For EBLM~J1741$+$31 we also set a priors on the eclipse depth $L$ and on the log of the stellar density $log \rho$, which is directly related to the transit parameters via Kepler's law \citep{pycheops}.
    This target has no detectable secondary eclipse and the primary eclipse is very shallow so the model parameters are poorly constrained by the light curve alone.
    The prior on eclipse depth was set using the predicted flux ratio.
    This ratio was calculated using the predicted absolute G-band magnitude, \textit{$M_G$}, for each star based on their  masses using the calibration by \citet{mamajek}.
    The scatter around the \textit{$M_G$}-mass relation for M-dwarfs was assumed to be similar to the observed scatter in \textit{$M_V$} magnitude values reported by \cite{hartman2015hats}.
    The prior for $log \rho$ was calculated using the derived values of mass and radius described in Section~\ref{sec:obs}.
    The values used for these priors are shown in Table \ref{Priors}.
    
\begin{figure*}
    \centering
    \begin{subfigure}{\textwidth}
        \centering
        \includegraphics[width=0.99\linewidth,height=0.4\linewidth]{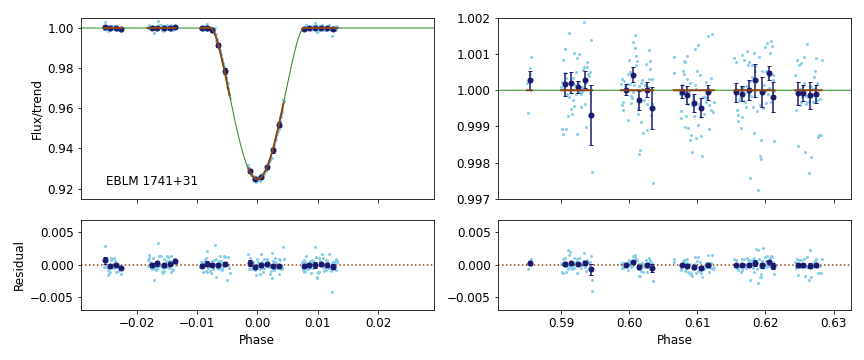} 
        \label{fig:subim1}
    \end{subfigure}
    \hfill
    \begin{subfigure}{\textwidth}
        \centering
        \includegraphics[width=0.99\linewidth,height=0.4\linewidth]{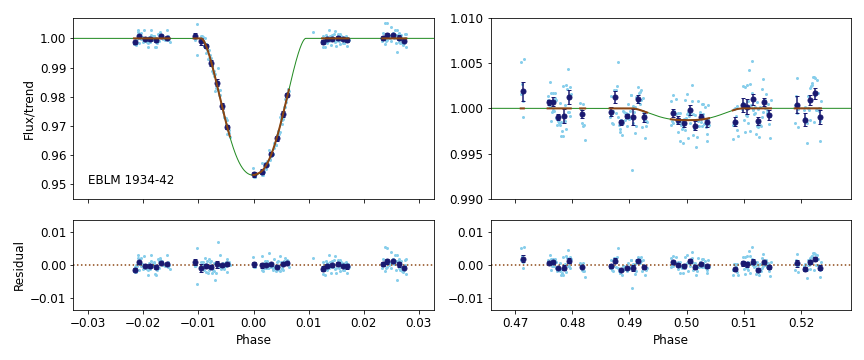}
        \label{fig:subim2}
    \end{subfigure}
    \hfill
    \begin{subfigure}{\textwidth}
        \centering
        \includegraphics[width=0.99\linewidth,height=0.4\linewidth]{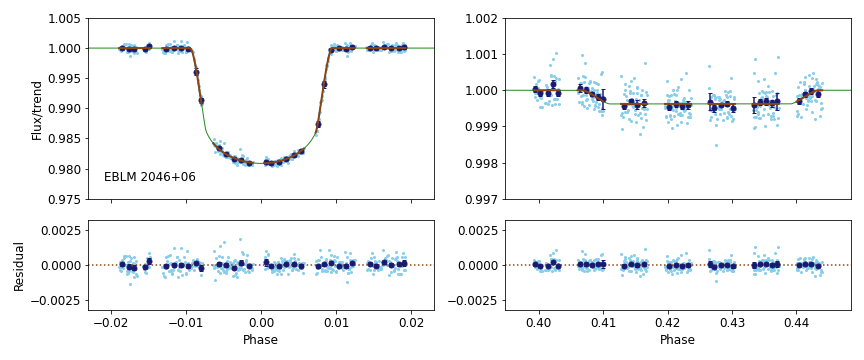}
        \label{fig:subim3}
    \end{subfigure}
    \caption{Fitted light curve of EBLM~J1741+31 (\textit{Top}), EBLM~J1934$-$42 (\textit{Middle}) and EBLM~J2046+06 (\textit{Bottom}) in phase intervals around the transit and eclipse events.
    The observed data corrected for instrumental trends according to the decorrelation coefficients given in Table~\ref{decorcoeff} are shown in cyan.
    The transit and eclipse models are shown in green.
    Binned data points with error bars are shown in blue and the fit between binned data points in brown.
    The residual of the fit is displayed below the fitted curves.}
    \label{fig:figure1}
\end{figure*}

    The joint posterior probability distribution (PPD) for the model and nuisance parameters are sampled using the sampler {\fontfamily{qcr}\selectfont emcee} \citep{Foreman}.
    The initial parameters of the run were the values previously obtained by the fits to the individual visit.
    We sampled a chain of 128 walkers each going through 35\,000 steps after a  ``burn-in'' of 1024 steps to ensure that the sampler has converged to a steady state.
    To ensure adequate sampling we ensured that the number of steps chosen was more than 50 times longer than the auto-correlation length of each fitted parameter chain.
    For EBLM~J1934$-$42 this required a second run of {\fontfamily{qcr}\selectfont emcee} with 180\,000 steps, and for EBLM~J1741+31 a second run with 240\,000 steps with a ``burn-in'' of 8192 steps.
    To ensure independent random samples from their posterior probability distributions, each parameter chain was thinned by approximately half the minimum parameter auto-correlation length.
    The parameter values given in Table~\ref{Params} are the median value of the parameters from the sampled PPD and the standard errors are estimated from the 15.9\% and 89.1\% percentile-points in the distribution for each parameter.
    The fitted decorrelation parameters from our analyses are shown in Appendix \ref{sec:decorr} in Table \ref{decorcoeff}.
    Correlations between selected parameters are displayed in Appendix \ref{sec:corners}.
    In EBLM~J1741+31 there are very strong correlations between $D$, $W$ and $b$ as can be seen in Figure \ref{fig:corner1741_cheops_combined}.
    In EBLM~J1934$-$42 the correlation between these parameters is not as strong though there are a significant number of walkers that tend to larger values of $D$ and $b$ as can be seen in Figure \ref{fig:corner1934_cheops_combined}.
    In EBLM~J2046+06 as shown in Figure \ref{fig:corner2046_cheops_combined} there is again a correlation between $D$, $W$ and $b$, but not as strongly as for EBLM~J1741+31.
    The light curve fit and residuals for these parameter values are shown in Fig.~\ref{fig:figure1}.

\subsection{TESS light curve analysis}

    We have compared our results using CHEOPS data to a similar analysis of the TESS light curves for EBLM~J1741+31 and EBLM~J1934$-$42. For EBLM~J1741+31 we used data from TESS sectors 25 and 26 covering 5 transits.
    For EBLM~J1934$-$42 we used data from sectors 13 and 27 covering 6 transits.
    Sampling of the posterior probability distribution of our model parameters was again performed using {\fontfamily{qcr}\selectfont emcee}.
    Gaussian priors were set on $f_c$ and $f_s$ using the same spectroscopically derived values as in the \textit{CHEOPS} fit. 
    Gaussian priors were also set on $h_1$ and $h_2$ using the stellar parameters given in Section~\ref{sec:obs} and assuming the same offset, but using the \textit{TESS} passband to interpolate our values.
    For EBLM~J1741+31 a prior on eclipse depth $L$ was again set using the predicted flux ratio of the target, adjusting to \textit{$M_{\rm Ic}$} magnitudes from \cite{mamajek} due to the different passband of \textit{TESS}.
    EBLM~J1741+31 required more steps than EBLM~J1934$-$42 to ensure the number of steps in the simulation was more than 50 times longer than the autocorrelation length in each parameter chain.
    We sampled a chain of 128 walkers each going through 20480 steps for EBLM~J1934$-$42 and 81920 steps for EBLM~J1741+31, with initial orbital parameters determined by a least-squares fit of the light curves.
    To allow the walkers to settle into the probability distributions we performed a burn-in of 2560 and 5120 steps before the sampling for EBLM~J1934$-$42 and EBLM~J1741+31, respectively.
    The parameter values given in Table~\ref{Params} are the median value of the parameters from the sampled PPD and the standard errors are estimated from the 15.9\% and 89.1\% percentile-points in the distribution for each parameter.
    In EBLM~J1741+31, similarly to the \textit{CHEOPS} light curve, there are very strong correlations between $D$, $W$ and $b$ as can be seen in Figure \ref{fig:corner1741_tess_combined}.
    In EBLM~J1934$-$42 the correlation between these parameters is not as strong. 
    Though there are a small amount of walkers that tend to larger values of $D$ and $b$ as can be seen in Figure \ref{fig:corner1934_tess_combined}, this is a smaller trend than in the \textit{CHEOPS} light curve.
    The light curve fit and residuals are shown in Fig.~\ref{fig:figure2}.

\subsection{Mass, radius and effective temperature}

    To obtain values of companion mass and radius we made use of the function \textsc{massradius} in {\fontfamily{qcr}\selectfont pycheops}.
    The M-dwarf mass is determined from the assumed primary mass $M_1$, orbital period $P$, orbital eccentricity $e$, the sine of orbital inclination $\sin(i)$ and the semi-amplitude of the star's spectroscopic orbit $K$.
    The M-dwarf radius is determines from the primary star radius $R_1$ from Table~\ref{ParamsStar} and the planet-star radius ratio from the light curve analysis, $k$.
    The value of $\log g_2$ in Table~\ref{Params} is determined directly from $K$ and the parameters of the transit light curve using equations (4) from \citet{2007MNRAS.379L..11S}.

    The ratio of the eclipse depths is directly related to the surface brightness ratio, i.e. $F_2/F_1 = L/D$, where $F_2$ is the flux per unit area integrated of the observing bandpass for star 2, and similarly for $F_1$.
    The surface brightness is directly related to a star's effective temperature, so we can use this information together with the values of $T_{\rm eff, 1}$, $\log g_1$ and [Fe/H] from Table~\ref{ParamsStar}, and spectral energy distributions from model stellar atmospheres to determine $T_{\rm eff, 2}$, the effective temperature of the M-dwarf.
    We calculated integrated surface brightness values for a large range of effective temperature, surface gravity and metallicity using PHOENIX model atmospheres with no alpha-element enhancement \citep{Husser} for both the \textit{CHEOPS} or \textit{TESS} bandpasses.
    We then sample the PPD for $T_{\rm eff, 2}$ using {\fontfamily{qcr}\selectfont emcee} and interpolation within this grid using the value of $\log g_2$ from Table~\ref{Params}.
    The results are given in Table~\ref{Params}.

\subsection{J1741+31 Eclipse Visit}

\begin{figure*}
    \centering
    \begin{subfigure}{\textwidth}
        \centering
        \includegraphics[height=0.4\linewidth]{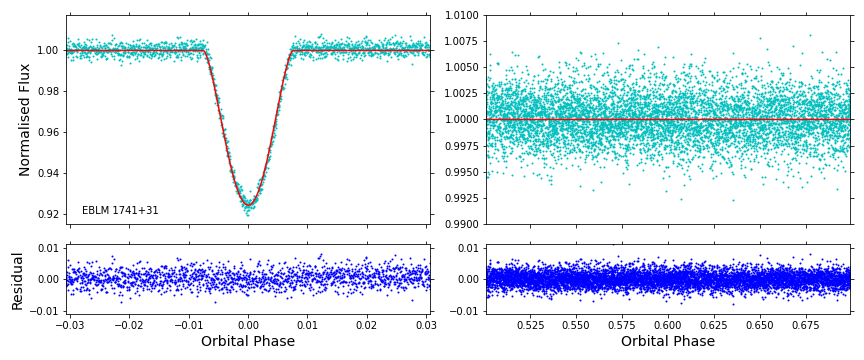} 
        \label{fig:subim4}
    \end{subfigure}
    \hfill
    \begin{subfigure}{\textwidth}
        \centering
        \includegraphics[height=0.4\linewidth]{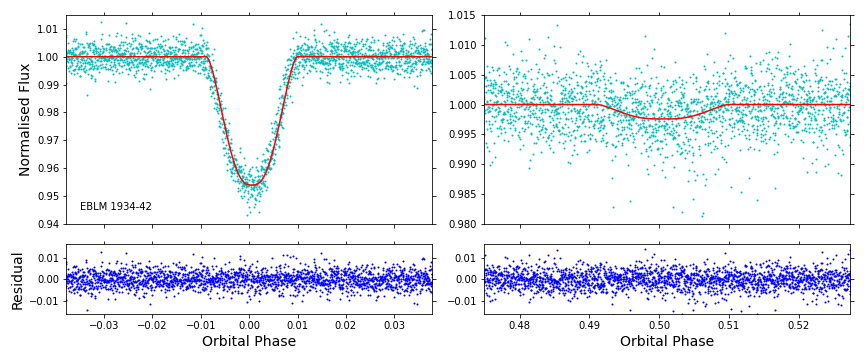}
        \label{fig:subim5}
    \end{subfigure}
    \caption{Fitted \textit{TESS} light curve of EBLM~J1741+31 (\textit{Top}) and EBLM~J1934$-$42 (\textit{Bottom}) in phase intervals around the transit and eclipse events.
    The observed data points are shown in cyan.
    The fitted light curve is shown in red.
    The residual of the fit is displayed below the fitted curves in blue.}
    \label{fig:figure2}
\end{figure*}
    
    Unfortunately, there is no secondary eclipse visible in the {\it CHEOPS} light curve for EBLM~J1741+31.
    We found that the predicted time of superior conjunction for our fitted model parameters is outside the duration of our scheduled {\it CHEOPS} visit.
    This visit was scheduled based on a preparatory analysis using less data than is now available for this target.
    We can use the analysis of the transit in the {\it CHEOPS} light curve to calculate the minimum separation of the stars around superior conjunction.
    We find that the probability that there is a secondary eclipse is $<0.002\%$.
    This explains why there is also no secondary eclipse visible in the \textit{TESS} light curve (Fig.~\ref{fig:figure2}).

\begin{table*}
    \centering
      \caption{The derived orbital parameters for each \textit{CHEOPS} target calculated by our {\fontfamily{qcr}\selectfont pycheops} fit.
      The eclipse depths displayed are in the relevant instrumental bandpass.}
         \label{Params}
    $$ 
         \begin{array}{lrrrrr}
            \hline
            \hline
            \noalign{\smallskip}
            \text{}      & \multicolumn{2}{c}{\text{J1741+31}} & \multicolumn{2}{c}{\text{J1934-42}} & \multicolumn{1}{c}{\text{J2046+06}}\\
             \text{}    & \multicolumn{1}{c}{\textit{CHEOPS}} & \multicolumn{1}{c}{\textit{TESS}} & \multicolumn{1}{c}{\textit{CHEOPS}} & \multicolumn{1}{c}{\textit{TESS}} & \multicolumn{1}{c}{\textit{CHEOPS}} \\
            \noalign{\smallskip}
            \hline
            \text{} {\rm Model~parameters} \\
            \hline            \hline
            T_0 \, ({\rm BJD}) & 2014.0490 \pm 0.0001 & 1990.9112 \pm 0.0001 & 2028.2295 \pm 0.0002 & 1659.7836 \pm 0.0002 & 2090.6246 \pm 0.0001 \\
            P \, ({\rm days}) & =7.71263 & 7.71263 \pm 0.00004 & =6.35251 & 6.35251 \pm 0.00001 & =10.10779 \\
            D & 0.152 \pm 0.024 & 0.109 \pm 0.011 & 0.0513 \pm 0.0047 & 0.0485 \pm 0.0011 & 0.0161 \pm 0.0002 \\
            W & 0.0091 \pm 0.0016 & 0.0118 \pm 0.0008 & 0.0190 \pm 0.0002 & 0.0189 \pm 0.0001 & 0.0263 \pm 0.0002 \\
            b & 1.312 \pm 0.061 & 1.184 \pm 0.041 & 0.797 \pm 0.027 & 0.785 \pm 0.009 & 0.165 \pm 0.096 \\
            f_c & 0.3006 \pm 0.0016\rlap{$^\dagger$} & 0.3003 \pm 0.0015\rlap{$^\dagger$} & =0.0 & =0.0 & -0.1902 \pm 0.0006\rlap{$^\dagger$} \\
            f_s & 0.4595 \pm 0.0012\rlap{$^\dagger$} & 0.4590 \pm 0.0012\rlap{$^\dagger$} & =0.0 & =0.0 & 0.5545 \pm 0.0004\rlap{$^\dagger$}\\
            L & -- & -- & 0.00126 \pm 0.00032 & 0.00250 \pm 0.00019 & 0.00039 \pm 0.00005   \\
            h_1 & 0.768 \pm 0.012\rlap{$^\dagger$}  & 0.818 \pm 0.011\rlap{$^\dagger$} & 0.729 \pm 0.011\rlap{$^\dagger$} & 0.784 \pm 0.011\rlap{$^\dagger$} & 0.757 \pm 0.011 \\
            h_2 & 0.435 \pm 0.050\rlap{$^\dagger$} & 0.397 \pm 0.050\rlap{$^\dagger$} & 0.398 \pm 0.050\rlap{$^\dagger$} & 0.394 \pm 0.050\rlap{$^\dagger$} & 0.393 \pm 0.178 \\
            \hline
            \noalign{\smallskip}
            \text{} {\rm Derived~parameters} \\
            \hline
            \noalign{\smallskip}
            R_2/R_1 & 0.390 \pm 0.031 & 0.330 \pm 0.017 & 0.2266 \pm 0.0102 & 0.2202 \pm 0.0025 & 0.1268 \pm 0.0007 \\
            R_1/a & 0.0621 \pm 0.0003 & 0.0610 \pm 0.0004 & 0.0639 \pm 0.0014 & 0.0634 \pm 0.0007 & 0.0743 \pm 0.0005      \\
            R_2/a & 0.0224 \pm 0.0019 & 0.0191 \pm 0.0011 & 0.0139 \pm 0.0010 & 0.0137 \pm 0.0003 & 0.0094 \pm 0.0001   \\
            i \, \text{ ($^{\circ}$)} & 85.32 \pm 0.22 & 85.86 \pm 0.17 & 87.08 \pm 0.16 & 87.15 \pm 0.06 & 89.30 \pm 0.41 \\
            e & 0.3015 \pm 0.0015  & 0.3009 \pm 0.0015 & 0.0 & 0.0 & 0.3437 \pm 0.0005 \\
            \omega \, \text{ ($^{\circ}$)} & 56.81 \pm 0.16 & 56.81 \pm 0.16 & -- & -- & 108.93 \pm 0.06\\
            \noalign{\smallskip}
            \hline
            \text{} {\rm Absolute~parameters} \\
            \hline
            \noalign{\smallskip}
            {\rm M}_2 \, ({\rm M_\odot}) & 0.4786 \pm 0.0095 & 0.4783 \pm 0.0095 & 0.1864 \pm 0.0055 & 0.1864 \pm 0.0055 & 0.1975 \pm 0.0053 \\
            {\rm R}_2 \, ({\rm R_\odot}) & 0.521 \pm 0.042 & 0.441 \pm 0.023 & 0.226 \pm 0.010 & 0.2193 \pm 0.0031 & 0.2184 \pm 0.0023 \\
            \log g_2 \, ({\rm cgs}) & 4.758 \pm 0.069 & 4.917 \pm 0.046 & 5.008 \pm 0.043 & 5.039 \pm 0.014 & 5.073 \pm 0.008\\
            {\rm T}_{\rm eff,2} \, ({\rm K}) &--&--& 3023 \pm 96 & 3030 \pm 41 & 3199 \pm 57 \\
            \hline
         \end{array}
    $$
    \medskip
    \parbox{\textwidth}{$^\dagger$: Derived parameters based on Gaussian priors shown in Table \ref{Priors}. 
     
    }
\end{table*}

\section{Discussion}
\label{sec:discuss}

    Observations of EBLM systems with \textit{CHEOPS} are complementary to the data provided by the \textit{TESS} mission.
    The \textit{CHEOPS} instrument response extends much further to the blue than \textit{TESS}.
    Looking for consistency of the transit parameters measured by the two instruments makes it possible to check for colour-dependent systematic errors, e.g. contamination of the photometry by other stars in the line of sight.
    Our results for EBLM~J1741+31 and EBLM~J1934$-$42 show good agreement between the results from the analysis of the \textit{CHEOPS}  and \textit{TESS} light curves.
    \textit{CHEOPS} is also able to observe regions of the sky not covered by the \textit{TESS} survey, e.g. close to the ecliptic.
    The precision of the parameters derived per transit from each instrument are similar so the final radius measurement from the \textit{TESS} data will be more precise in cases where it has observed many transits.
    \textit{CHEOPS} observations can be scheduled to cover individual transit or eclipse events, which can be advantageous if we want to observe long-period systems. 
    
    Our results for EBLM~J2046+06 show that \textit{CHEOPS} light curves can be used to measure radii accurate to about 1\% and T$_{\rm eff}$ accurate to about 2\% for the M-dwarf in EBLM systems with well-defined transits.
    This is sufficient for our main goal of establishing an empirical mass-radius-metallicity relation for very low mass stars.
    Observations of 24 additional EBLM binaries with well-defined transits with \textit{CHEOPS} are on-going. 
    The results presented here have already been used by \citet{pycheops} to constrain the properties of the host star in their study of the super-Earth GJ 1132\,b using \textit{CHEOPS} observations of the transit.
    
    The transit model in {\fontfamily{qcr}\selectfont pycheops} does not account for surface features on the primary star due to magnetic activity, e.g. dark spots, faculae or plages.
    The impact of these features on the parameters derived is dependent on whether they are occulted by the secondary star or not \citep{czesla2009stellar, pont2013prevalence, 2013A+A...556A..19O}.
    Dark spots occulted during the transit will produce small peaks in the light curve.
    If these are not accounted for in the model then the transit depth will be underestimated, leading to an underestimate for the companion radius.
    The opposite is true for dark spots not occulted by the companion. We checked the TESS and WASP light curves of our targets for variability on time scales of a few days or more due to the combination of rotation and magnetic activity.
    For all three targets we find that any such variability has an amplitude $\loa 1$\% ($\loa 0.1$\% for EBLM~J2046+06).
    Spots near the poles of these slowly-rotating solar-type stars are not expected so we conclude that magnetic activity has a negligible impact on the parameters we have derived for the M-dwarfs in these systems.
    
\begin{figure*}
    \centering
    \begin{subfigure}{0.5\textwidth}
        \centering
        \includegraphics[width=0.99\linewidth]{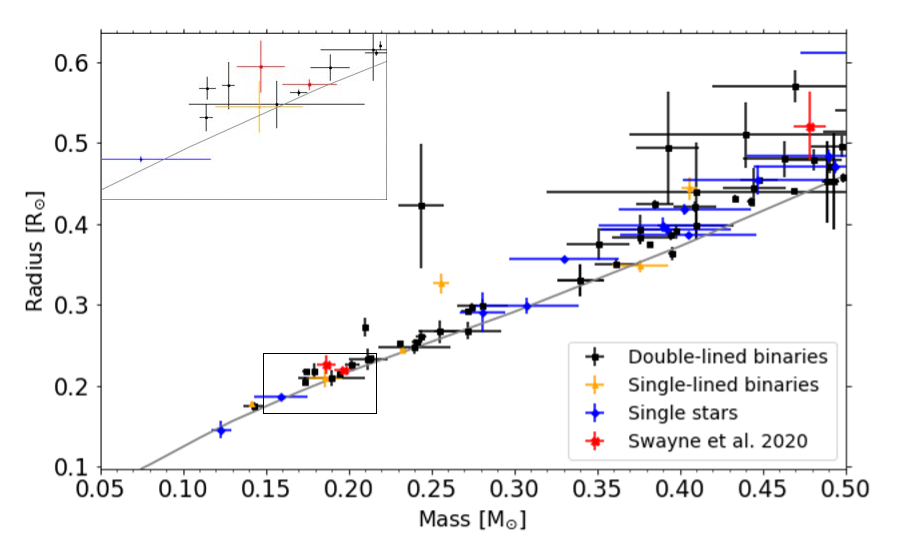} 
        \label{fig:subim6}
    \end{subfigure}\begin{subfigure}{0.5\textwidth}
        \centering
        \includegraphics[width=0.99\linewidth]{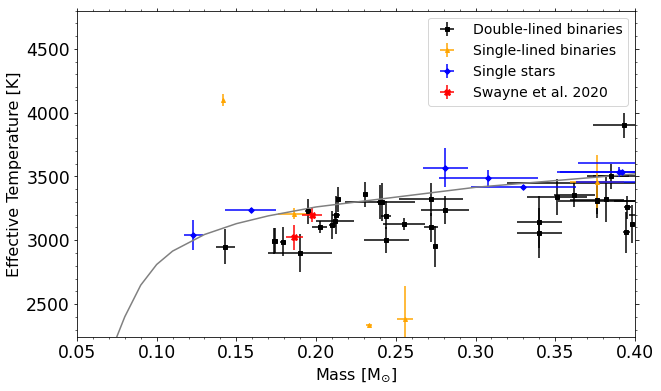}
        \label{fig:subim7}
    \end{subfigure}
    \caption{\textit{Left}: A cutout of the stellar mass versus stellar radius diagram using results from \citet{nefs2013,gillen,parsons2018scatter} with our results highlighted in red. 
    The type of system is displayed by different colours. 
    The theoretical relation from \citet{Baraffe15} for an age of 1 Gyr is plotted in gray.
    \textit{Right}: A cutout of the stellar mass versus effective temperature diagram using results from \citet{nefs2013,gillen,parsons2018scatter}, with our results highlighted in red. 
    The type of system is displayed by different colours. 
    The theoretical relation from \citet{Baraffe15} for an age of 1 Gyr is plotted in gray.}
    \label{fig:figure3}
\end{figure*}

    Our results are shown in the context of other mass, radius and effective temperature measurements for M-dwarfs in Fig.~\ref{fig:figure3}.
    EBLM~J1741+31 and EBLM~J1934$-$42 follow the trend for stars with masses $\loa 0.5$\,M$_{\odot}$ to be larger on average by a few percent than predicted by models that do not account for magnetic activity.
    The radius of EBLM~J2046+06, which is our most precise radius measurement, agrees well with the models of \citet{Baraffe15}.
    EBLM~J1934$-$42 is a metal-rich star, which may be consistent with the idea that metallicity has an influence on radius inflation \cite[e.g.][]{berger2006, spada2013, von2019eblm}.
    Not shown in Fig.~\ref{fig:figure3} are the masses and radii for M-dwarfs in EBLM binaries by \citet{von2019eblm} and \citet{gill2019eblm}.
    We do not yet have effective temperature measurements for these M-dwarfs, but the methods we have developed here can be applied to the {\it CHEOPS} and {\it TESS} light curves for those stars, as well as other EBLM binaries observed by these instruments, to provide a more complete picture for these systems. 

\section{Conclusions}
\label{sec:conc}
    In this paper we have reported the first results of our \textit{CHEOPS} observing programme on low-mass eclipsing binaries.
    We find that the very high precision of the photometry from this instrument and the possibility to schedule observations of individual transit and eclipse events are well-matched to our science goal of measuring an empirical mass-radius-metallicity relation for very low mass stars.
    We report three M-dwarf radii and two effective temperatures between our three targets contributing to the rather sparse amount of data at the low-mass end of the H-R diagram.
    Additional observations from our on-going observations with \textit{CHEOPS} complemented by further analysis of data from the \textit{TESS} mission will provide precise and accurate mass, radius and T$_{\rm eff}$ measurements for many very low-mass stars of known metallicity and age.
    Fundamental data of this quality will be essential if we are to find an answer to the long-standing radius inflation problem.  
    
\section*{Acknowledgements}

CHEOPS is an ESA mission in partnership with Switzerland with important contributions to the payload and the ground segment from Austria, Belgium, France, Germany, Hungary, Italy, Portugal, Spain, Sweden, and the United Kingdom. The CHEOPS Consortium would like to gratefully acknowledge the support received by all the agencies, offices, universities, and industries involved. Their flexibility and willingness to explore new approaches were essential to the success of this mission.
This publication makes use of The Data \& Analysis Center for Exoplanets (DACE), which is a facility based at the University of Geneva (CH) dedicated to extrasolar planets data visualisation, exchange and analysis. 
DACE is a platform of the Swiss National Centre of Competence in Research (NCCR) PlanetS, federating the Swiss expertise in Exoplanet research. 
The DACE platform is available at \url{https://dace.unige.ch}.
MIS thanks Giovanni Bruno for his comments on the draft of this manuscript.
MIS and PFLM are supported by the UK Science and Technology Facilities Council (STFC) grant numbers ST/M001040/1 and ST/T506175/1.
This research is supported from the European Research Council (ERC) under the European Union's Horizon 2020 research and innovation programme (grant agreement n$^\circ$ 803193/BEBOP), by a Leverhulme Trust Research Project Grant (n$^\circ$ RPG-2018-418), and by observations obtained at the Observatoire de Haute-Provence (CNRS), France (PI Santerne).
S.G.S. acknowledge support from FCT through FCT contract nr. CEECIND/00826/2018 and POPH/FSE (EC).
IB, AS and TL acknowledge financial support from the French Programme National de Plan{\'e}tologie (PNP, INSU).
Support for this work was provided by NASA through the NASA Hubble Fellowship grant \#HST-HF2-51464 awarded by the Space Telescope Science Institute, which is operated by the Association of Universities for Research in Astronomy, Inc., for NASA, under contract NAS5-26555.
S.S. has received funding from the European Research Council (ERC) under the European Union’s Horizon 2020 research and innovation programme (grant agreement No 833925, project STAREX).
V.V.G. is an F.R.S-FNRS Research Associate. 
ACC and TW acknowledge support from STFC consolidated grant number ST/M001296/1.
YA and MJH  acknowledge  the  support  of  the  Swiss  National  Fund  under  grant 200020\_172746. 
We acknowledge support from the Spanish Ministry of Science and Innovation and the European Regional Development Fund through grants ESP2016-80435-C2-1-R, ESP2016-80435-C2-2-R, PGC2018-098153-B-C33, PGC2018-098153-B-C31, ESP2017-87676-C5-1-R, MDM-2017-0737 Unidad de Excelencia ``Mar\'{\i}a de Maeztu''- Centro de Astrobiología (INTA-CSIC), as well as the support of the Generalitat de Catalunya/CERCA programme. The MOC activities have been supported by the ESA contract No. 4000124370. 
S.C.C.B. acknowledges support from FCT through FCT contracts nr. IF/01312/2014/CP1215/CT0004. 
XB, SC, DG, MF and JL acknowledge their role as ESA-appointed CHEOPS science team members. 
ABr was supported by the SNSA.
This project was supported by the CNES.
LD is an F.R.S.-FNRS Postdoctoral Researcher.
The Belgian participation to CHEOPS has been supported by the Belgian Federal Science Policy Office (BELSPO) in the framework of the PRODEX Program, and by the University of Li{\`e}ge through an ARC grant for Concerted Research Actions financed by the Wallonia-Brussels Federation. 
This work was supported by FCT - Funda\c{c}\~{a}o para a Ci\^{e}ncia e a Tecnologia through national funds and by FEDER through COMPETE2020 - Programa Operacional Competitividade e Internacionaliza\c{c}\~{a}o by these grants: UID/FIS/04434/2019; UIDB/04434/2020; UIDP/04434/2020; PTDC/FIS-AST/32113/2017 \& POCI-01-0145-FEDER- 032113; PTDC/FIS-AST/28953/2017 \& POCI-01-0145-FEDER-028953; PTDC/FIS-AST/28987/2017 \& POCI-01-0145-FEDER-028987.
O.D.S.D. is supported in the form of work contract (DL 57/2016/CP1364/CT0004) funded by national funds through FCT. 
B.-O.D. acknowledges support from the Swiss National Science Foundation (PP00P2-190080). 
This project has received funding from the European Research Council (ERC) under the European Union's Horizon 2020 research and innovation programme (project {\sc Four Aces}; grant agreement No 724427). 
This project has been carried out in the frame of the National Centre for Competence in Research PlanetS supported by the Swiss National Science Foundation (SNSF). 
MF and CMP gratefully acknowledge the support of the Swedish National Space Agency (DNR 65/19, 174/18). 
DG gratefully acknowledges financial support from the CRT foundation under Grant No. 2018.2323 ``Gaseous or rocky? Unveiling the nature of small worlds''. 
M.G. is an F.R.S.-FNRS Senior Research Associate. 
MH acknowledges support from ANID – Millennium Science Initiative – ICN12 009.
SH gratefully acknowledges CNES funding through the grant 837319.
KGI is the ESA CHEOPS Project Scientist and is responsible for the ESA CHEOPS Guest Observers Programme. She does not participate in, or contribute to, the definition of the Guaranteed Time Programme of the CHEOPS mission through which observations described in this paper have been taken, nor to any aspect of target selection for the programme.
This work was granted access to the HPC resources of MesoPSL financed by the Region Ile de France and the project Equip@Meso (reference ANR-10-EQPX-29-01) of the programme Investissements d'Avenir supervised by the Agence Nationale pour la Recherche. 
This work was also partially supported by a grant from the Simons Foundation (PI Queloz, grant number 327127).
This  project  has  been  supported  by  the  Hungarian National Research, Development and Innovation Office (NKFIH) grants GINOP-2.3.2-15-2016-00003, K-119517,  K-125015, and the City of Szombathely under Agreement No.\ 67.177-21/2016.

\section*{Data Availability}
The CHEOPS data underlying this article are available in its online supplementary material.
The fitted light curve is available at CDS via anonymous ftp to \url{cdsarc.u-strasbg.fr} (130.79.128.5) or via \url{https://cdsarc.unistra.fr/viz-bin/cat/J/MNRAS}.

This paper includes data collected by the TESS mission, which is publicly available from the Mikulski Archive for Space Telescopes (MAST) at the Space Telescope Science Institure (STScI) (\url{https://mast.stsci.edu}). 
Funding for the TESS mission is provided by the NASA Explorer Program directorate. 
STScI is operated by the Association of Universities for Research in Astronomy, Inc., under NASA contract NAS 5–26555.
We acknowledge the use of public TESS Alert data from pipelines at the TESS Science Office and at the TESS Science Processing Operations Center.
 


\bibliographystyle{mnras}
\bibliography{0mybib}

\begin{thebibliography}{}
\makeatletter
\relax
\def\mn@urlcharsother{\let\do\@makeother \do\$\do\&\do\#\do\^\do\_\do\%\do\~}
\def\mn@doi{\begingroup\mn@urlcharsother \@ifnextchar [ {\mn@doi@}
  {\mn@doi@[]}}
\def\mn@doi@[#1]#2{\def\@tempa{#1}\ifx\@tempa\@empty \href
  {http://dx.doi.org/#2} {doi:#2}\else \href {http://dx.doi.org/#2} {#1}\fi
  \endgroup}
\def\mn@eprint#1#2{\mn@eprint@#1:#2::\@nil}
\def\mn@eprint@arXiv#1{\href {http://arxiv.org/abs/#1} {{\tt arXiv:#1}}}
\def\mn@eprint@dblp#1{\href {http://dblp.uni-trier.de/rec/bibtex/#1.xml}
  {dblp:#1}}
\def\mn@eprint@#1:#2:#3:#4\@nil{\def\@tempa {#1}\def\@tempb {#2}\def\@tempc
  {#3}\ifx \@tempc \@empty \let \@tempc \@tempb \let \@tempb \@tempa \fi \ifx
  \@tempb \@empty \def\@tempb {arXiv}\fi \@ifundefined
  {mn@eprint@\@tempb}{\@tempb:\@tempc}{\expandafter \expandafter \csname
  mn@eprint@\@tempb\endcsname \expandafter{\@tempc}}}

\bibitem[\protect\citeauthoryear{{Baraffe}, {Chabrier}, {Allard}  \&
  {Hauschildt}}{{Baraffe} et~al.}{1998}]{Baraffe98}
{Baraffe} I.,  {Chabrier} G.,  {Allard} F.,   {Hauschildt} P.~H.,  1998, \aap,
  \href {https://ui.adsabs.harvard.edu/abs/1998A&A...337..403B} {337, 403}

\bibitem[\protect\citeauthoryear{{Baraffe}, {Homeier}, {Allard}  \&
  {Chabrier}}{{Baraffe} et~al.}{2015}]{Baraffe15}
{Baraffe} I.,  {Homeier} D.,  {Allard} F.,   {Chabrier} G.,  2015, \mn@doi
  [\aap] {10.1051/0004-6361/201425481}, \href
  {https://ui.adsabs.harvard.edu/abs/2015A&A...577A..42B} {577, A42}

\bibitem[\protect\citeauthoryear{{Baranne} et~al.,}{{Baranne}
  et~al.}{1996}]{Baranne96}
{Baranne} A.,  et~al., 1996, \aaps, \href
  {https://ui.adsabs.harvard.edu/abs/1996A&AS..119..373B} {119, 373}

\bibitem[\protect\citeauthoryear{{Benz} et~al.,}{{Benz} et~al.}{2021}]{Benz}
{Benz} W.,  et~al., 2021, \mn@doi [Experimental Astronomy]
  {10.1007/s10686-020-09679-4}, \href
  {https://ui.adsabs.harvard.edu/abs/2021ExA....51..109B} {51, 109}

\bibitem[\protect\citeauthoryear{{Berger} et~al.,}{{Berger}
  et~al.}{2006}]{berger2006}
{Berger} D.~H.,  et~al., 2006, \mn@doi [\apj] {10.1086/503318}, \href
  {https://ui.adsabs.harvard.edu/abs/2006ApJ...644..475B} {644, 475}

\bibitem[\protect\citeauthoryear{{Blackwell} \& {Shallis}}{{Blackwell} \&
  {Shallis}}{1977}]{Blackwell1977}
{Blackwell} D.~E.,  {Shallis} M.~J.,  1977, \mn@doi [\mnras]
  {10.1093/mnras/180.2.177}, \href
  {https://ui.adsabs.harvard.edu/abs/1977MNRAS.180..177B} {180, 177}

\bibitem[\protect\citeauthoryear{{Bonfanti}, {Ortolani}, {Piotto}  \&
  {Nascimbeni}}{{Bonfanti} et~al.}{2015}]{bonfanti15}
{Bonfanti} A.,  {Ortolani} S.,  {Piotto} G.,   {Nascimbeni} V.,  2015, \mn@doi
  [\aap] {10.1051/0004-6361/201424951}, \href
  {http://adsabs.harvard.edu/abs/2015A%26A...575A..18B} {575, A18}

\bibitem[\protect\citeauthoryear{{Bonfanti}, {Ortolani}  \&
  {Nascimbeni}}{{Bonfanti} et~al.}{2016}]{bonfanti16}
{Bonfanti} A.,  {Ortolani} S.,   {Nascimbeni} V.,  2016, \mn@doi [\aap]
  {10.1051/0004-6361/201527297}, \href
  {http://adsabs.harvard.edu/abs/2016A%26A...585A...5B} {585, A5}

\bibitem[\protect\citeauthoryear{{Bonfanti} et~al.,}{{Bonfanti}
  et~al.}{2021}]{bonfanti21}
{Bonfanti} A.,  et~al., 2021, \mn@doi [\aap] {10.1051/0004-6361/202039608},
  \href {https://ui.adsabs.harvard.edu/abs/2021A&A...646A.157B} {646, A157}

\bibitem[\protect\citeauthoryear{{Bouchy}, {D{\'\i}az}, {H{\'e}brard},
  {Arnold}, {Boisse}, {Delfosse}, {Perruchot}  \& {Santerne}}{{Bouchy}
  et~al.}{2013}]{Bouchy2013}
{Bouchy} F.,  {D{\'\i}az} R.~F.,  {H{\'e}brard} G.,  {Arnold} L.,  {Boisse} I.,
   {Delfosse} X.,  {Perruchot} S.,   {Santerne} A.,  2013, \mn@doi [\aap]
  {10.1051/0004-6361/201219979}, \href
  {https://ui.adsabs.harvard.edu/abs/2013A&A...549A..49B} {549, A49}

\bibitem[\protect\citeauthoryear{{Casagrande}, {Flynn}  \&
  {Bessell}}{{Casagrande} et~al.}{2008}]{casagrande2008}
{Casagrande} L.,  {Flynn} C.,   {Bessell} M.,  2008, \mn@doi [\mnras]
  {10.1111/j.1365-2966.2008.13573.x}, \href
  {https://ui.adsabs.harvard.edu/abs/2008MNRAS.389..585C} {389, 585}

\bibitem[\protect\citeauthoryear{{Castelli} \& {Kurucz}}{{Castelli} \&
  {Kurucz}}{2003}]{Castelli2003}
{Castelli} F.,  {Kurucz} R.~L.,  2003, in {Piskunov} N.,  {Weiss} W.~W.,
  {Gray} D.~F.,  eds,  IAU Symposium Vol. 210, Modelling of Stellar
  Atmospheres. p.~A20 (\mn@eprint {arXiv} {astro-ph/0405087})

\bibitem[\protect\citeauthoryear{{Chabrier}, {Gallardo}  \&
  {Baraffe}}{{Chabrier} et~al.}{2007}]{chabrier2007evolution}
{Chabrier} G.,  {Gallardo} J.,   {Baraffe} I.,  2007, \mn@doi [\aap]
  {10.1051/0004-6361:20077702}, \href
  {https://ui.adsabs.harvard.edu/abs/2007A&A...472L..17C} {472, L17}

\bibitem[\protect\citeauthoryear{{Charbonneau} \& {Deming}}{{Charbonneau} \&
  {Deming}}{2007}]{charbonneau2007dynamics}
{Charbonneau} D.,  {Deming} D.,  2007, arXiv e-prints, \href
  {https://ui.adsabs.harvard.edu/abs/2007arXiv0706.1047C} {p. arXiv:0706.1047}

\bibitem[\protect\citeauthoryear{{Courcol} et~al.,}{{Courcol}
  et~al.}{2015}]{Courcol15}
{Courcol} B.,  et~al., 2015, \mn@doi [\aap] {10.1051/0004-6361/201526329},
  \href {https://ui.adsabs.harvard.edu/abs/2015A&A...581A..38C} {581, A38}

\bibitem[\protect\citeauthoryear{{Czesla}, {Huber}, {Wolter}, {Schr{\"o}ter}
  \& {Schmitt}}{{Czesla} et~al.}{2009}]{czesla2009stellar}
{Czesla} S.,  {Huber} K.~F.,  {Wolter} U.,  {Schr{\"o}ter} S.,   {Schmitt}
  J.~H.~M.~M.,  2009, \mn@doi [\aap] {10.1051/0004-6361/200912454}, \href
  {https://ui.adsabs.harvard.edu/abs/2009A&A...505.1277C} {505, 1277}

\bibitem[\protect\citeauthoryear{{Delrez} et~al.,}{{Delrez}
  et~al.}{2018}]{delrez}
{Delrez} L.,  et~al., 2018, in {Marshall} H.~K.,  {Spyromilio} J.,  eds,
  Society of Photo-Optical Instrumentation Engineers (SPIE) Conference Series
  Vol. 10700, Ground-based and Airborne Telescopes VII. p. 107001I (\mn@eprint
  {arXiv} {1806.11205}), \mn@doi{10.1117/12.2312475}

\bibitem[\protect\citeauthoryear{{Dotter}, {Chaboyer}, {Jevremovi{\'c}},
  {Kostov}, {Baron}  \& {Ferguson}}{{Dotter} et~al.}{2008}]{Dotter}
{Dotter} A.,  {Chaboyer} B.,  {Jevremovi{\'c}} D.,  {Kostov} V.,  {Baron} E.,
  {Ferguson} J.~W.,  2008, \mn@doi [\apjs] {10.1086/589654}, \href
  {https://ui.adsabs.harvard.edu/abs/2008ApJS..178...89D} {178, 89}

\bibitem[\protect\citeauthoryear{{Feiden} \& {Chaboyer}}{{Feiden} \&
  {Chaboyer}}{2014}]{feiden2014revised}
{Feiden} G.~A.,  {Chaboyer} B.,  2014, \mn@doi [\aap]
  {10.1051/0004-6361/201424288}, \href
  {https://ui.adsabs.harvard.edu/abs/2014A&A...571A..70F} {571, A70}

\bibitem[\protect\citeauthoryear{{Fernandes}, {Van Grootel}, {Salmon},
  {Aringer}, {Burgasser}, {Scuflaire}, {Brassard}  \& {Fontaine}}{{Fernandes}
  et~al.}{2019}]{fernandes2019evolutionary}
{Fernandes} C.~S.,  {Van Grootel} V.,  {Salmon} S. J.~A.~J.,  {Aringer} B.,
  {Burgasser} A.~J.,  {Scuflaire} R.,  {Brassard} P.,   {Fontaine} G.,  2019,
  \mn@doi [\apj] {10.3847/1538-4357/ab2333}, \href
  {https://ui.adsabs.harvard.edu/abs/2019ApJ...879...94F} {879, 94}

\bibitem[\protect\citeauthoryear{{Foreman-Mackey}, {Hogg}, {Lang}  \&
  {Goodman}}{{Foreman-Mackey} et~al.}{2013}]{Foreman}
{Foreman-Mackey} D.,  {Hogg} D.~W.,  {Lang} D.,   {Goodman} J.,  2013, \mn@doi
  [\pasp] {10.1086/670067}, \href
  {https://ui.adsabs.harvard.edu/abs/2013PASP..125..306F} {125, 306}

\bibitem[\protect\citeauthoryear{{Foreman-Mackey}, {Agol}, {Ambikasaran}  \&
  {Angus}}{{Foreman-Mackey} et~al.}{2017}]{celerite}
{Foreman-Mackey} D.,  {Agol} E.,  {Ambikasaran} S.,   {Angus} R.,  2017,
  \mn@doi [\aj] {10.3847/1538-3881/aa9332}, \href
  {https://ui.adsabs.harvard.edu/abs/2017AJ....154..220F} {154, 220}

\bibitem[\protect\citeauthoryear{{Gaia Collaboration} et~al.,}{{Gaia
  Collaboration} et~al.}{2018}]{gaia2018gaia}
{Gaia Collaboration} et~al., 2018, \mn@doi [\aap]
  {10.1051/0004-6361/201833051}, \href
  {https://ui.adsabs.harvard.edu/abs/2018A&A...616A...1G} {616, A1}

\bibitem[\protect\citeauthoryear{{Gaia Collaboration}, {Brown}, {Vallenari},
  {Prusti}, {de Bruijne}, {Babusiaux}  \& {Biermann}}{{Gaia Collaboration}
  et~al.}{2020}]{GaiaCollaboration2020}
{Gaia Collaboration} {Brown} A.~G.~A.,  {Vallenari} A.,  {Prusti} T.,  {de
  Bruijne} J.~H.~J.,  {Babusiaux} C.,   {Biermann} M.,  2020, arXiv e-prints,
  \href {https://ui.adsabs.harvard.edu/abs/2020arXiv201201533G} {p.
  arXiv:2012.01533}

\bibitem[\protect\citeauthoryear{{Gill} et~al.,}{{Gill}
  et~al.}{2019}]{gill2019eblm}
{Gill} S.,  et~al., 2019, \mn@doi [\aap] {10.1051/0004-6361/201833054}, \href
  {https://ui.adsabs.harvard.edu/abs/2019A&A...626A.119G} {626, A119}

\bibitem[\protect\citeauthoryear{{Gillen}, {Hillenbrand}, {David}, {Aigrain},
  {Rebull}, {Stauffer}, {Cody}  \& {Queloz}}{{Gillen} et~al.}{2017}]{gillen}
{Gillen} E.,  {Hillenbrand} L.~A.,  {David} T.~J.,  {Aigrain} S.,  {Rebull} L.,
   {Stauffer} J.,  {Cody} A.~M.,   {Queloz} D.,  2017, \mn@doi [\apj]
  {10.3847/1538-4357/aa84b3}, \href
  {https://ui.adsabs.harvard.edu/abs/2017ApJ...849...11G} {849, 11}

\bibitem[\protect\citeauthoryear{{Gillon} et~al.,}{{Gillon}
  et~al.}{2017}]{gillon2017seven}
{Gillon} M.,  et~al., 2017, \mn@doi [\nat] {10.1038/nature21360}, \href
  {https://ui.adsabs.harvard.edu/abs/2017Natur.542..456G} {542, 456}

\bibitem[\protect\citeauthoryear{{G{\'o}mez Maqueo Chew} et~al.,}{{G{\'o}mez
  Maqueo Chew} et~al.}{2014}]{Yilen}
{G{\'o}mez Maqueo Chew} Y.,  et~al., 2014, \mn@doi [\aap]
  {10.1051/0004-6361/201424265}, \href
  {https://ui.adsabs.harvard.edu/abs/2014A&A...572A..50G} {572, A50}

\bibitem[\protect\citeauthoryear{{Hara} et~al.,}{{Hara} et~al.}{2020}]{Hara20}
{Hara} N.~C.,  et~al., 2020, \mn@doi [\aap] {10.1051/0004-6361/201937254},
  \href {https://ui.adsabs.harvard.edu/abs/2020A&A...636L...6H} {636, L6}

\bibitem[\protect\citeauthoryear{{Hartman} et~al.,}{{Hartman}
  et~al.}{2015}]{hartman2015hats}
{Hartman} J.~D.,  et~al., 2015, \mn@doi [\aj] {10.1088/0004-6256/149/5/166},
  \href {https://ui.adsabs.harvard.edu/abs/2015AJ....149..166H} {149, 166}

\bibitem[\protect\citeauthoryear{{Hoyer}, {Guterman}, {Demangeon}, {Sousa},
  {Deleuil}, {Meunier}  \& {Benz}}{{Hoyer} et~al.}{2020}]{hoyer2020expected}
{Hoyer} S.,  {Guterman} P.,  {Demangeon} O.,  {Sousa} S.~G.,  {Deleuil} M.,
  {Meunier} J.~C.,   {Benz} W.,  2020, \mn@doi [\aap]
  {10.1051/0004-6361/201936325}, \href
  {https://ui.adsabs.harvard.edu/abs/2020A&A...635A..24H} {635, A24}

\bibitem[\protect\citeauthoryear{{Husser}, {Wende-von Berg}, {Dreizler},
  {Homeier}, {Reiners}, {Barman}  \& {Hauschildt}}{{Husser}
  et~al.}{2013}]{Husser}
{Husser} T.~O.,  {Wende-von Berg} S.,  {Dreizler} S.,  {Homeier} D.,  {Reiners}
  A.,  {Barman} T.,   {Hauschildt} P.~H.,  2013, \mn@doi [\aap]
  {10.1051/0004-6361/201219058}, \href
  {https://ui.adsabs.harvard.edu/abs/2013A&A...553A...6H} {553, A6}

\bibitem[\protect\citeauthoryear{{Irwin} et~al.,}{{Irwin}
  et~al.}{2011}]{irwin2011lspm}
{Irwin} J.~M.,  et~al., 2011, \mn@doi [\apj] {10.1088/0004-637X/742/2/123},
  \href {https://ui.adsabs.harvard.edu/abs/2011ApJ...742..123I} {742, 123}

\bibitem[\protect\citeauthoryear{{Jenkins} et~al.,}{{Jenkins}
  et~al.}{2016}]{jenkins2016tess}
{Jenkins} J.~M.,  et~al., 2016, in {Chiozzi} G.,  {Guzman} J.~C.,  eds,
  Society of Photo-Optical Instrumentation Engineers (SPIE) Conference Series
  Vol. 9913, Software and Cyberinfrastructure for Astronomy IV. p. 99133E,
  \mn@doi{10.1117/12.2233418}

\bibitem[\protect\citeauthoryear{{Kesseli}, {Muirhead}, {Mann}  \&
  {Mace}}{{Kesseli} et~al.}{2018}]{kesseli2018}
{Kesseli} A.~Y.,  {Muirhead} P.~S.,  {Mann} A.~W.,   {Mace} G.,  2018, \mn@doi
  [\aj] {10.3847/1538-3881/aabccb}, \href
  {https://ui.adsabs.harvard.edu/abs/2018AJ....155..225K} {155, 225}

\bibitem[\protect\citeauthoryear{{Kurucz}}{{Kurucz}}{1993}]{Kurucz-93}
{Kurucz} R.~L.,  1993, {SYNTHE spectrum synthesis programs and line data}.
Cambridge, MA: Smithsonian Astrophysical Observatory

\bibitem[\protect\citeauthoryear{{Lindegren} et~al.,}{{Lindegren}
  et~al.}{2020}]{Lindegren2020}
{Lindegren} L.,  et~al., 2020, arXiv e-prints, \href
  {https://ui.adsabs.harvard.edu/abs/2020arXiv201201742L} {p. arXiv:2012.01742}

\bibitem[\protect\citeauthoryear{{Magic}, {Chiavassa}, {Collet}  \&
  {Asplund}}{{Magic} et~al.}{2015}]{magic2015stagger}
{Magic} Z.,  {Chiavassa} A.,  {Collet} R.,   {Asplund} M.,  2015, \mn@doi
  [\aap] {10.1051/0004-6361/201423804}, \href
  {https://ui.adsabs.harvard.edu/abs/2015A&A...573A..90M} {573, A90}

\bibitem[\protect\citeauthoryear{{Marigo} et~al.,}{{Marigo}
  et~al.}{2017}]{marigo17}
{Marigo} P.,  et~al., 2017, \mn@doi [\apj] {10.3847/1538-4357/835/1/77}, \href
  {http://adsabs.harvard.edu/abs/2017ApJ...835...77M} {835, 77}

\bibitem[\protect\citeauthoryear{{Martin} et~al.,}{{Martin}
  et~al.}{2019}]{martin2019bebop}
{Martin} D.~V.,  et~al., 2019, \mn@doi [\aap] {10.1051/0004-6361/201833669},
  \href {https://ui.adsabs.harvard.edu/abs/2019A&A...624A..68M} {624, A68}

\bibitem[\protect\citeauthoryear{{Maxted}}{{Maxted}}{2016}]{maxted2016ellc}
{Maxted} P.~F.~L.,  2016, \mn@doi [\aap] {10.1051/0004-6361/201628579}, \href
  {https://ui.adsabs.harvard.edu/abs/2016A&A...591A.111M} {591, A111}

\bibitem[\protect\citeauthoryear{{Maxted}}{{Maxted}}{2018}]{maxted2018}
{Maxted} P.~F.~L.,  2018, \mn@doi [\aap] {10.1051/0004-6361/201832944}, \href
  {https://ui.adsabs.harvard.edu/abs/2018A&A...616A..39M} {616, A39}

\bibitem[\protect\citeauthoryear{{Maxted} \& {Gill}}{{Maxted} \&
  {Gill}}{2019}]{maxted2019q}
{Maxted} P.~F.~L.,  {Gill} S.,  2019, \mn@doi [\aap]
  {10.1051/0004-6361/201834563}, \href
  {https://ui.adsabs.harvard.edu/abs/2019A&A...622A..33M} {622, A33}

\bibitem[\protect\citeauthoryear{{Maxted} et~al.}{{Maxted}
  et~al.}{2021}]{pycheops}
{Maxted} P. F.~L.,  et~al., 2021, \mnras, submitted.

\bibitem[\protect\citeauthoryear{{Morales} et~al.,}{{Morales}
  et~al.}{2009}]{morales2009absolute}
{Morales} J.~C.,  et~al., 2009, \mn@doi [\apj] {10.1088/0004-637X/691/2/1400},
  \href {https://ui.adsabs.harvard.edu/abs/2009ApJ...691.1400M} {691, 1400}

\bibitem[\protect\citeauthoryear{{Nefs} et~al.,}{{Nefs}
  et~al.}{2013}]{nefs2013}
{Nefs} S.~V.,  et~al., 2013, \mn@doi [\mnras] {10.1093/mnras/stt405}, \href
  {https://ui.adsabs.harvard.edu/abs/2013MNRAS.431.3240N} {431, 3240}

\bibitem[\protect\citeauthoryear{{Olander}, {Heiter}  \& {Kochukhov}}{{Olander}
  et~al.}{2021}]{olander2021comparative}
{Olander} T.,  {Heiter} U.,   {Kochukhov} O.,  2021, arXiv e-prints, \href
  {https://ui.adsabs.harvard.edu/abs/2021arXiv210208836O} {p. arXiv:2102.08836}

\bibitem[\protect\citeauthoryear{{Oshagh}, {Santos}, {Boisse}, {Bou{\'e}},
  {Montalto}, {Dumusque}  \& {Haghighipour}}{{Oshagh}
  et~al.}{2013}]{2013A+A...556A..19O}
{Oshagh} M.,  {Santos} N.~C.,  {Boisse} I.,  {Bou{\'e}} G.,  {Montalto} M.,
  {Dumusque} X.,   {Haghighipour} N.,  2013, \mn@doi [\aap]
  {10.1051/0004-6361/201321309}, \href
  {https://ui.adsabs.harvard.edu/abs/2013A&A...556A..19O} {556, A19}

\bibitem[\protect\citeauthoryear{{Parsons} et~al.,}{{Parsons}
  et~al.}{2018}]{parsons2018scatter}
{Parsons} S.~G.,  et~al., 2018, \mn@doi [\mnras] {10.1093/mnras/sty2345}, \href
  {https://ui.adsabs.harvard.edu/abs/2018MNRAS.481.1083P} {481, 1083}

\bibitem[\protect\citeauthoryear{{Pecaut} \& {Mamajek}}{{Pecaut} \&
  {Mamajek}}{2013}]{mamajek}
{Pecaut} M.~J.,  {Mamajek} E.~E.,  2013, \mn@doi [\apjs]
  {10.1088/0067-0049/208/1/9}, \href
  {https://ui.adsabs.harvard.edu/abs/2013ApJS..208....9P} {208, 9}

\bibitem[\protect\citeauthoryear{{Perruchot} et~al.,}{{Perruchot}
  et~al.}{2008}]{Perruchot08}
{Perruchot} S.,  et~al., 2008, in {McLean} I.~S.,  {Casali} M.~M.,  eds,
  Society of Photo-Optical Instrumentation Engineers (SPIE) Conference Series
  Vol. 7014, Ground-based and Airborne Instrumentation for Astronomy II. p.
  70140J, \mn@doi{10.1117/12.787379}

\bibitem[\protect\citeauthoryear{{Pollacco} et~al.,}{{Pollacco}
  et~al.}{2006}]{pollacco2006wasp}
{Pollacco} D.~L.,  et~al., 2006, \mn@doi [\pasp] {10.1086/508556}, \href
  {https://ui.adsabs.harvard.edu/abs/2006PASP..118.1407P} {118, 1407}

\bibitem[\protect\citeauthoryear{{Pont}, {Sing}, {Gibson}, {Aigrain}, {Henry}
  \& {Husnoo}}{{Pont} et~al.}{2013}]{pont2013prevalence}
{Pont} F.,  {Sing} D.~K.,  {Gibson} N.~P.,  {Aigrain} S.,  {Henry} G.,
  {Husnoo} N.,  2013, \mn@doi [\mnras] {10.1093/mnras/stt651}, \href
  {https://ui.adsabs.harvard.edu/abs/2013MNRAS.432.2917P} {432, 2917}

\bibitem[\protect\citeauthoryear{{Quirrenbach} et~al.,}{{Quirrenbach}
  et~al.}{2014}]{quirrenbach2014carmenes}
{Quirrenbach} A.,  et~al., 2014, in {Ramsay} S.~K.,  {McLean} I.~S.,   {Takami}
  H.,  eds,  Society of Photo-Optical Instrumentation Engineers (SPIE)
  Conference Series Vol. 9147, Ground-based and Airborne Instrumentation for
  Astronomy V. p. 91471F, \mn@doi{10.1117/12.2056453}

\bibitem[\protect\citeauthoryear{{Ribas}}{{Ribas}}{2006}]{ribas2006}
{Ribas} I.,  2006, \mn@doi [\apss] {10.1007/s10509-006-9081-4}, \href
  {https://ui.adsabs.harvard.edu/abs/2006Ap&SS.304...89R} {304, 89}

\bibitem[\protect\citeauthoryear{{Ricker} et~al.,}{{Ricker}
  et~al.}{2015}]{Ricker}
{Ricker} G.~R.,  et~al., 2015, \mn@doi [Journal of Astronomical Telescopes,
  Instruments, and Systems] {10.1117/1.JATIS.1.1.014003}, \href
  {https://ui.adsabs.harvard.edu/abs/2015JATIS...1a4003R} {1, 014003}

\bibitem[\protect\citeauthoryear{{Salmon}, {Van Grootel}, {Buldgen}, {Dupret}
  \& {Eggenberger}}{{Salmon} et~al.}{2021}]{salmon21}
{Salmon} S.~J.~A.~J.,  {Van Grootel} V.,  {Buldgen} G.,  {Dupret} M.~A.,
  {Eggenberger} P.,  2021, \mn@doi [\aap] {10.1051/0004-6361/201937174}, \href
  {https://ui.adsabs.harvard.edu/abs/2021A&A...646A...7S} {646, A7}

\bibitem[\protect\citeauthoryear{{Santos} et~al.,}{{Santos}
  et~al.}{2013}]{Santos-13}
{Santos} N.~C.,  et~al., 2013, \mn@doi [\aap] {10.1051/0004-6361/201321286},
  \href {http://adsabs.harvard.edu/abs/2013A%26A...556A.150S} {556, A150}

\bibitem[\protect\citeauthoryear{{Santos}, {Barros}, {Demangeon}  \&
  {Faria}}{{Santos} et~al.}{2020}]{santos2020detection}
{Santos} N.~C.,  {Barros} S. C.~C.,  {Demangeon} O. D.~S.,   {Faria} J.~P.,
  2020, {Detection and Characterization Methods of Exoplanets}.
Oxford University Press, p.~189,
  \mn@doi{10.1093/acrefore/9780190647926.013.189}

\bibitem[\protect\citeauthoryear{{Schanche} et~al.,}{{Schanche}
  et~al.}{2020}]{Schanche2020}
{Schanche} N.,  et~al., 2020, \mn@doi [\mnras] {10.1093/mnras/staa2848}, \href
  {https://ui.adsabs.harvard.edu/abs/2020MNRAS.499..428S} {499, 428}

\bibitem[\protect\citeauthoryear{{Scuflaire}, {Th{\'e}ado}, {Montalb{\'a}n},
  {Miglio}, {Bourge}, {Godart}, {Thoul}  \& {Noels}}{{Scuflaire}
  et~al.}{2008}]{scuflaire08}
{Scuflaire} R.,  {Th{\'e}ado} S.,  {Montalb{\'a}n} J.,  {Miglio} A.,  {Bourge}
  P.-O.,  {Godart} M.,  {Thoul} A.,   {Noels} A.,  2008, \mn@doi [\apss]
  {10.1007/s10509-007-9650-1}, \href
  {http://adsabs.harvard.edu/abs/2008Ap%26SS.316...83S} {316, 83}

\bibitem[\protect\citeauthoryear{{Skrutskie} et~al.,}{{Skrutskie}
  et~al.}{2006}]{Skrutskie2006}
{Skrutskie} M.~F.,  et~al., 2006, \mn@doi [\aj] {10.1086/498708}, \href
  {https://ui.adsabs.harvard.edu/abs/2006AJ....131.1163S} {131, 1163}

\bibitem[\protect\citeauthoryear{{Sneden}}{{Sneden}}{1973}]{Sneden-73}
{Sneden} C.~A.,  1973, PhD thesis, THE UNIVERSITY OF TEXAS AT AUSTIN.

\bibitem[\protect\citeauthoryear{{Sousa}}{{Sousa}}{2014}]{Sousa-14}
{Sousa} S.~G.,  2014, {ARES + MOOG: A Practical Overview of an Equivalent Width
  (EW) Method to Derive Stellar Parameters}.
Springer International Publishing (Cham), pp 297--310,
  \mn@doi{10.1007/978-3-319-06956-2_26}

\bibitem[\protect\citeauthoryear{{Sousa}, {Santos}, {Israelian}, {Mayor}  \&
  {Monteiro}}{{Sousa} et~al.}{2007}]{Sousa-07}
{Sousa} S.~G.,  {Santos} N.~C.,  {Israelian} G.,  {Mayor} M.,   {Monteiro}
  M.~J.~P.~F.~G.,  2007, \mn@doi [A\&A] {10.1051/0004-6361:20077288}, \href
  {http://adsabs.harvard.edu/abs/2007A%26A...469..783S} {469, 783}

\bibitem[\protect\citeauthoryear{{Sousa} et~al.,}{{Sousa}
  et~al.}{2008}]{Sousa-08}
{Sousa} S.~G.,  et~al., 2008, \mn@doi [A\&A] {10.1051/0004-6361:200809698},
  \href {http://adsabs.harvard.edu/abs/2008A%26A...487..373S} {487, 373}

\bibitem[\protect\citeauthoryear{{Sousa}, {Santos}, {Adibekyan}, {Delgado-Mena}
   \& {Israelian}}{{Sousa} et~al.}{2015}]{Sousa-15}
{Sousa} S.~G.,  {Santos} N.~C.,  {Adibekyan} V.,  {Delgado-Mena} E.,
  {Israelian} G.,  2015, \mn@doi [\aap] {10.1051/0004-6361/201425463}, \href
  {http://adsabs.harvard.edu/abs/2015A%26A...577A..67S} {577, A67}

\bibitem[\protect\citeauthoryear{{Southworth}, {Wheatley}  \&
  {Sams}}{{Southworth} et~al.}{2007}]{2007MNRAS.379L..11S}
{Southworth} J.,  {Wheatley} P.~J.,   {Sams} G.,  2007, \mn@doi [\mnras]
  {10.1111/j.1745-3933.2007.00324.x}, \href
  {https://ui.adsabs.harvard.edu/abs/2007MNRAS.379L..11S} {379, L11}

\bibitem[\protect\citeauthoryear{{Spada}, {Demarque}, {Kim}  \&
  {Sills}}{{Spada} et~al.}{2013}]{spada2013}
{Spada} F.,  {Demarque} P.,  {Kim} Y.~C.,   {Sills} A.,  2013, \mn@doi [\apj]
  {10.1088/0004-637X/776/2/87}, \href
  {https://ui.adsabs.harvard.edu/abs/2013ApJ...776...87S} {776, 87}

\bibitem[\protect\citeauthoryear{{Swayne}, {Maxted}, {Hod{\v{z}}i{\'c}}  \&
  {Triaud}}{{Swayne} et~al.}{2020}]{swayne2020tess}
{Swayne} M.~I.,  {Maxted} P. F.~L.,  {Hod{\v{z}}i{\'c}} V.~K.,   {Triaud} A.
  H.~M.~J.,  2020, \mn@doi [\mnras] {10.1093/mnrasl/slaa122}, \href
  {https://ui.adsabs.harvard.edu/abs/2020MNRAS.498L..15S} {498, L15}

\bibitem[\protect\citeauthoryear{Tenenbaum \& Jenkins}{Tenenbaum \&
  Jenkins}{2018}]{tenenbaum}
Tenenbaum P.,  Jenkins J.,  2018, Technical report, TESS Science Data Products
  Description Document.
EXP-TESS-ARC-ICD-0014 Rev D https://archive. stsci. edu/missions/tess/doc~…

\bibitem[\protect\citeauthoryear{{Tognelli}, {Prada Moroni}  \&
  {Degl'Innocenti}}{{Tognelli} et~al.}{2018}]{tognelli2018theoretical}
{Tognelli} E.,  {Prada Moroni} P.~G.,   {Degl'Innocenti} S.,  2018, \mn@doi
  [\mnras] {10.1093/mnras/sty195}, \href
  {https://ui.adsabs.harvard.edu/abs/2018MNRAS.476...27T} {476, 27}

\bibitem[\protect\citeauthoryear{{Torres}, {Andersen}  \&
  {Gim{\'e}nez}}{{Torres} et~al.}{2010}]{torres2010}
{Torres} G.,  {Andersen} J.,   {Gim{\'e}nez} A.,  2010, \mn@doi [\aapr]
  {10.1007/s00159-009-0025-1}, \href
  {https://ui.adsabs.harvard.edu/abs/2010A&ARv..18...67T} {18, 67}

\bibitem[\protect\citeauthoryear{{Triaud} et~al.,}{{Triaud}
  et~al.}{2013}]{triaud}
{Triaud} A.~H.~M.~J.,  et~al., 2013, \mn@doi [\aap]
  {10.1051/0004-6361/201219643}, \href
  {https://ui.adsabs.harvard.edu/abs/2013A&A...549A..18T} {549, A18}

\bibitem[\protect\citeauthoryear{{Triaud} et~al.,}{{Triaud}
  et~al.}{2017}]{triaud2017eblm}
{Triaud} A. H.~M.~J.,  et~al., 2017, \mn@doi [\aap]
  {10.1051/0004-6361/201730993}, \href
  {https://ui.adsabs.harvard.edu/abs/2017A&A...608A.129T} {608, A129}

\bibitem[\protect\citeauthoryear{{Van Grootel} et~al.,}{{Van Grootel}
  et~al.}{2018}]{van2018stellar}
{Van Grootel} V.,  et~al., 2018, \mn@doi [\apj] {10.3847/1538-4357/aaa023},
  \href {https://ui.adsabs.harvard.edu/abs/2018ApJ...853...30V} {853, 30}

\bibitem[\protect\citeauthoryear{{Wright} et~al.,}{{Wright}
  et~al.}{2010}]{Wright2010}
{Wright} E.~L.,  et~al., 2010, \mn@doi [\aj] {10.1088/0004-6256/140/6/1868},
  \href {https://ui.adsabs.harvard.edu/abs/2010AJ....140.1868W} {140, 1868}

\bibitem[\protect\citeauthoryear{{von Boetticher} et~al.,}{{von Boetticher}
  et~al.}{2019}]{von2019eblm}
{von Boetticher} A.,  et~al., 2019, \mn@doi [\aap]
  {10.1051/0004-6361/201834539}, \href
  {https://ui.adsabs.harvard.edu/abs/2019A&A...625A.150V} {625, A150}

\makeatother
\end{thebibliography}

\bigskip

\noindent
\hrulefill \\
$^{1}$ Astrophysics Group, Keele University, Staffordshire, ST5 5BG, United Kingdom\\
$^{2}$ School of Physics and Astronomy, University of Birmingham, Edgbaston, Birmingham B15 2TT, UK\\
$^{3}$ Instituto de Astrof\'isica e Ci\^encias do Espa\c{c}o, Universidade do Porto, CAUP, Rua das Estrelas, 4150-762 Porto, Portugal\\
$^{4}$ Physikalisches Institut, University of Bern, Gesellsschaftstrasse 6, 3012 Bern, Switzerland\\
$^{5}$ Center for Space and Habitability, Gesellsschaftstrasse 6, 3012 Bern, Switzerland\\
$^{6}$ Department of Astronomy, Stockholm University, AlbaNova University Center, 10691 Stockholm, Sweden\\
$^{7}$ Aix Marseille Univ, CNRS, CNES, LAM, Marseille, France\\
$^{8}$ Division Technique INSU, BP 330, 83507 La Seyne cedex, France\\
$^{9}$ Space Research Institute, Austrian Academy of Sciences, Schmiedlstrasse 6, A-8042 Graz, Austria\\
$^{10}$ Department of Astronomy, The Ohio State University, Columbus, OH 43210, USA\\
$^{11}$ Observatoire Astronomique de l'Universit\'e de Gen\`eve, Chemin Pegasi 51, Versoix, Switzerland\\
$^{12}$ Space sciences, Technologies and Astrophysics Research (STAR) Institute, Universit{\'e} de Li{\`e}ge, All{\'e}e du 6 Ao{\^u}t 19C, 4000 Li{\`e}ge, Belgium\\
$^{13}$ Centre for Exoplanet Science, SUPA School of Physics and Astronomy, University of St Andrews, North Haugh, St Andrews KY16 9SS, UK\\
$^{14}$ Instituto de Astrof\'\i sica de Canarias, 38200 La Laguna, Tenerife, Spain\\
$^{15}$ Departamento de Astrof\'\i sica, Universidad de La Laguna, 38206 La Laguna, Tenerife, Spain\\
$^{16}$ Institut de Ci\`encies de l'Espai (ICE, CSIC), Campus UAB, Can Magrans s/n, 08193 Bellaterra, Spain\\
$^{17}$ Institut d'Estudis Espacials de Catalunya (IEEC), 08034 Barcelona, Spain\\
$^{18}$ ESTEC, European Space Agency, 2201AZ, Noordwijk, NL\\
$^{19}$ Admatis, Miskok, Hungary\\
$^{20}$ Depto. de Astrofísica, Centro de Astrobiologia (CSIC-INTA), ESAC campus, 28692 Villanueva de la Cãda (Madrid), Spain\\
$^{21}$ Departamento de F\'isica e Astronomia, Faculdade de Ci\^encias, Universidade do Porto, Rua do Campo Alegre, 4169-007 Porto, Portugal\\
$^{22}$ Department of Physics, University of Warwick, Gibbet Hill Road, Coventry CV4 7AL, United Kingdom\\
$^{23}$ Centre for Exoplanets and Habitability, University of Warwick, Gibbet Hill Road, Coventry CV4 7AL, UK\\
$^{24}$ Université Grenoble Alpes, CNRS, IPAG, 38000 Grenoble, France\\
$^{25}$ Institute of Planetary Research, German Aerospace Center (DLR), Rutherfordstrasse 2, 12489 Berlin, Germany\\
$^{26}$ Université de Paris, Institut de physique du globe de Paris, CNRS, F-75005 Paris, France\\
$^{27}$ Lund Observatory, Dept. of Astronomy and Theoretical Physics, Lund University, Box 43, 22100 Lund, Sweden\\
$^{28}$ Astrobiology Research Unit, Universit\'e de Li\`ege, All\'ee du 6 Ao\^ut 19C, B-4000 Li\`ege, Belgium\\
$^{29}$ Leiden Observatory, University of Leiden, PO Box 9513, 2300 RA Leiden, The Netherlands\\
$^{30}$ Department of Space, Earth and Environment, Chalmers University of Technology, Onsala Space Observatory, 43992 Onsala, Sweden\\
$^{31}$ Dipartimento di Fisica, Universit\`a degli Studi di Torino, via Pietro Giuria 1, I-10125, Torino, Italy\\
$^{32}$ Department of Astrophysics, University of Vienna, Tuerkenschanzstrasse 17, 1180 Vienna, Austria\\
$^{33}$ Institut d'astrophysique de Paris, UMR7095 CNRS, Université Pierre \& Marie Curie, 98bis blvd. Arago, 75014 Paris, France\\
$^{34}$ Observatoire de Haute-Provence, CNRS, Universit\'e d'Aix-Marseille, 04870 Saint-Michel l'Observatoire, France\\
$^{35}$ Department of Physics, Shahid Beheshti University, Tehran, Iran.\\
$^{36}$ Laboratoire J.-L. Lagrange, Observatoire de la C{\^o}te d'Azur (OCA), Universite de Nice-Sophia Antipolis (UNS), CNRS, Campus Valrose, 06108 Nice Cedex 2, France\\
$^{37}$ Millennium Institute for Astrophysics, Chile\\
$^{38}$ Instituto de Astrofísica, Facultad de Física, Pontificia Universidad Cat\'olica de Chile, Av. Vicu\~{n}a Mackenna 4860, 782-0436 Macul, Santiago, Chile\\
$^{39}$ Science and Operations Department - Science Division (SCI-SC), Directorate of Science, European Space Agency (ESA), European Space Research and Technology Centre (ESTEC), Keplerlaan 1, 2201-AZ Noordwijk, The Netherlands\\
$^{40}$ Konkoly Observatory, Research Centre for Astronomy and Earth Sciences, 1121 Budapest, Konkoly Thege Miklós út 15-17, Hungary\\
$^{41}$ ELTE E\"otv\"os Lor\'and University, Institute of Physics, P\'azm\'any P\'eter s\'et\'any 1/A, 1117 Budapest, Hungary\\
$^{42}$ Sydney Institute for Astronomy, School of Physics A29, University of Sydney, NSW 2006, Australia\\
$^{43}$ IMCCE, UMR8028 CNRS, Observatoire de Paris, PSL Univ., Sorbonne Univ., 77 av. Denfert-Rochereau, 75014 Paris, France\\
$^{44}$ INAF, Osservatorio Astronomico di Padova, Vicolo dell'Osservatorio 5, 35122 Padova, Italy\\
$^{45}$ INAF, Osservatorio Astrofisico di Catania, Via S. Sofia 78, 95123 Catania, Italy\\
$^{46}$ Institute of Optical Sensor Systems, German Aerospace Center (DLR), Rutherfordstrasse 2, 12489 Berlin, Germany\\
$^{47}$ Dipartimento di Fisica e Astronomia "Galileo Galilei", Università degli Studi di Padova, Vicolo dell'Osservatorio 3, 35122 Padova, Italy\\
$^{48}$ Cavendish Laboratory, JJ Thomson Avenue, Cambridge CB3 0HE, UK\\
$^{49}$ Center for Astronomy and Astrophysics, Technical University Berlin, Hardenberstrasse 36, 10623 Berlin, Germany\\
$^{50}$ Institut für Geologische Wissenschaften, Freie Universität Berlin, 12249 Berlin, Germany\\
$^{51}$ ELTE Eötvös Loránd University, Gothard Astrophysical Observatory, 9700 Szombathely, Szent Imre h. u. 112, Hungary\\
$^{52}$ MTA-ELTE Exoplanet Research Group, 9700 Szombathely, Szent Imre h. u. 112, Hungary\\
$^{53}$ Institute of Astronomy, University of Cambridge, Madingley Road, Cambridge, CB3 0HA, United Kingdom\\



\appendix
\section{Decorrelation Parameters}
\label{sec:decorr}

\begin{table*}
\centering
      \caption{The decorrelation parameters fitted from the \textit{CHEOPS} MultiVisit MCMC analysis. The effects these parameters represent are as follows: image background level (dfdbg), PSF centroid position (dfdx, dfdy) time (dfdt), aperture contamination (dfdcontam) and  smear correction (dfdsmear).}
         \label{decorcoeff}
    
         \resizebox{\textwidth}{!}{\begin{tabular}{lcrrrrrr}
            \hline
            \hline
            \noalign{\smallskip}
            Target  & \multicolumn{1}{l}{Visit} & \multicolumn{1}{l}{dfdbg} & \multicolumn{1}{l}{dfdx} & \multicolumn{1}{l}{dfdy} & \multicolumn{1}{l}{dfdt} & \multicolumn{1}{l}{dfdcontam} & \multicolumn{1}{l}{dfdsmear} \\

               &   & \multicolumn{1}{l}{[$10^{-3}$]} & \multicolumn{1}{l}{$10^{-4}$} & \multicolumn{1}{l}{$10^{-3}$} & \multicolumn{1}{l}{[$10^{-2} {\rm d}^{-1}$]} & \multicolumn{1}{l}{$10^{-3}$} & \multicolumn{1}{l}{$10^{-4}$} \\

            \noalign{\smallskip}
            \hline
            \noalign{\smallskip}
            EBLM J1741+31 & Transit & -- & -- & -- & -- & -- & --  \\
             & Eclipse & -- & -- & --& -- & --  & --  \\
            EBLM J1934-42 & Transit & $-0.023 \pm 1.491$ &--& $-1.66 \pm 0.37$ &--& -- &--  \\
             & Eclipse &-- &--& $-0.38 \pm 0.33$ & $-1.04 \pm 0.13$ & $-3.25 \pm 0.79$ & -- \\
            EBLM 2046+06 & Transit & --&--&--&--& $-0.55 \pm 0.15$ & --   \\
             & Eclipse & -- & $-2.25 \pm 0.56$ & $0.40 \pm 0.05$ &--&-- & $9.45 \pm 1.72$ \\
            \noalign{\smallskip}
            \hline
         \end{tabular}}
    
\end{table*}

\section{Correlation Diagrams for Selected Parameters}
\label{sec:corners}

\begin{figure*}
\begin{centering}
\includegraphics[height=0.9\textwidth]{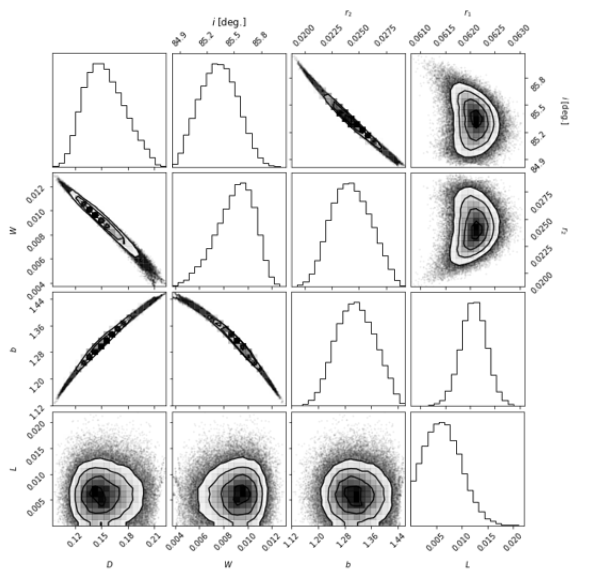} 
\end{centering}
\caption{Corner plot for \textit{CHEOPS} dataset of EBLM~J1741+31.}
\label{fig:corner1741_cheops_combined}
\end{figure*}

\begin{figure*}
\begin{centering}
\includegraphics[height=0.9\textwidth]{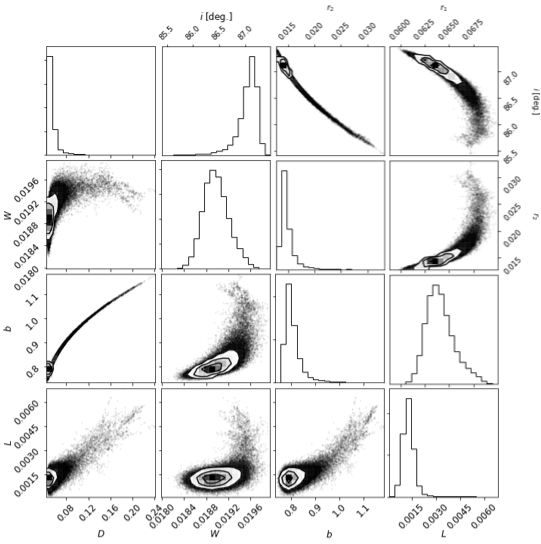} 
\end{centering}
\caption{Corner plot for \textit{CHEOPS} dataset of EBLM~J1934$-$42.}
\label{fig:corner1934_cheops_combined}
\end{figure*}

\begin{figure*}
\begin{centering}
\includegraphics[height=0.9\textwidth]{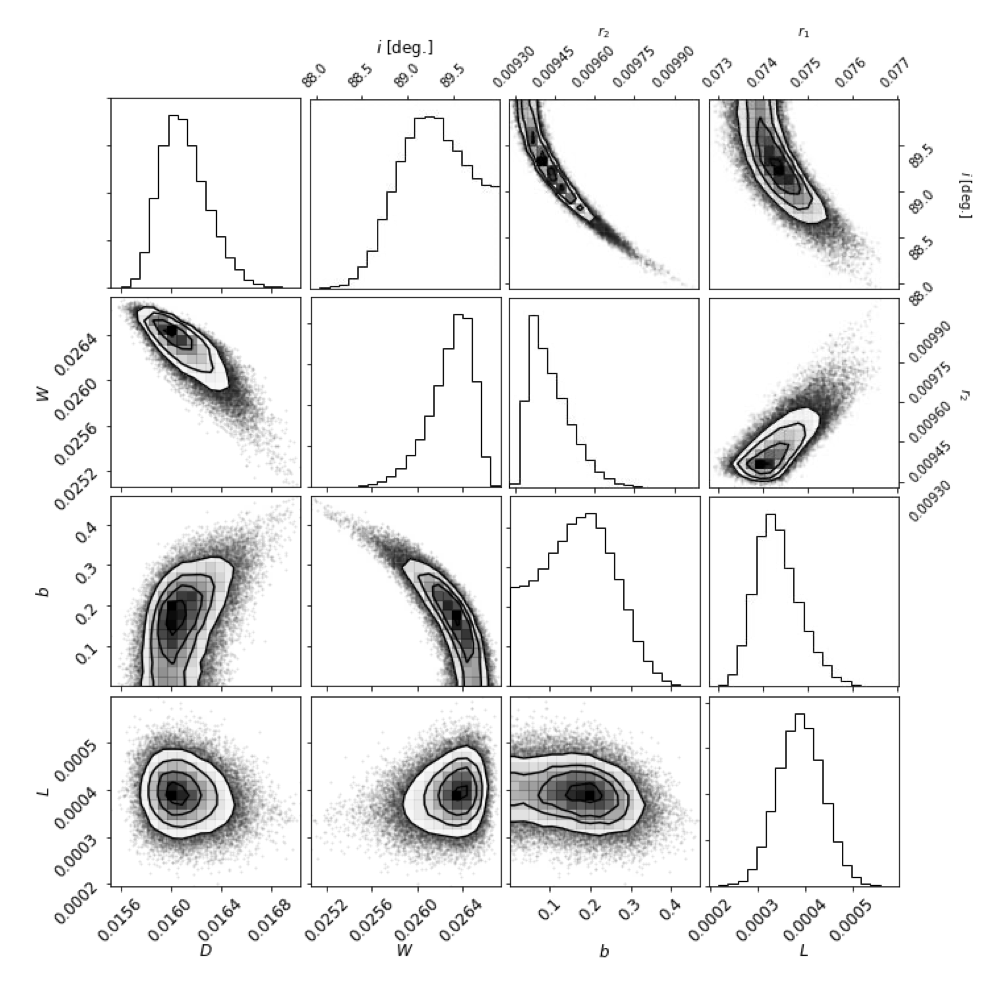} 
\end{centering}
\caption{Corner plot for \textit{CHEOPS} dataset of EBLM~J2046+06.}
\label{fig:corner2046_cheops_combined}
\end{figure*}

\begin{figure*}
\begin{centering}
\includegraphics[height=0.9\textwidth]{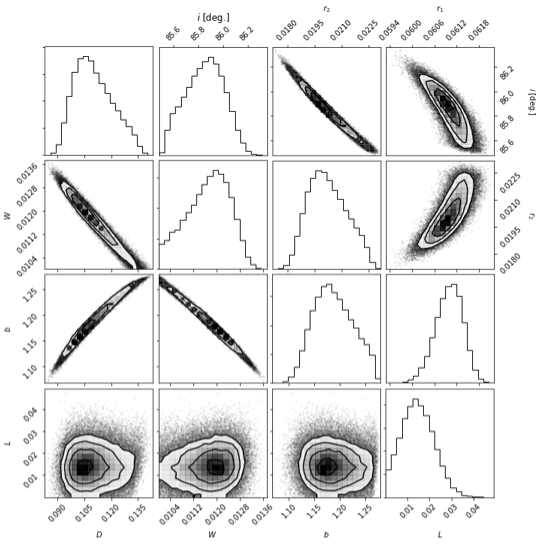} 
\end{centering}
\caption{Corner plot for \textit{TESS} dataset of EBLM~J1741+31.}
\label{fig:corner1741_tess_combined}
\end{figure*}

\begin{figure*}
\begin{centering}
\includegraphics[height=0.9\textwidth]{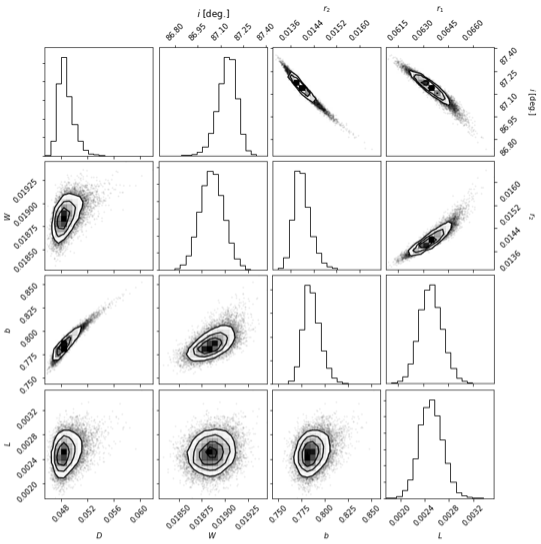} 
\end{centering}
\caption{Corner plot for \textit{TESS} dataset of EBLM~J1934$-$42.}
\label{fig:corner1934_tess_combined}
\end{figure*}
    
\bsp	
\label{lastpage}
\end{document}